\documentclass[aps,prl,twocolumn,superscriptaddress,
 nobibnotes,
 amsmath,amssymb,amsfonts,
 floatfix
]{revtex4-2}

\makeatletter
\def\p@subsection{}
\def\p@subsubsection{}
\makeatother

\pdfoutput=1
\usepackage{amsmath,amssymb}
\usepackage{graphicx}
\usepackage{color}
\usepackage{bm}
\usepackage{hyperref}
\hypersetup{
    colorlinks=true,
    citecolor=NavyBlue,
    linkcolor=BrickRed,
    urlcolor=NavyBlue,
    bookmarksopen=false,
}
\usepackage{float}
\usepackage[T1]{fontenc}
\usepackage{braket}
\usepackage{physics}
\usepackage{caption}
\usepackage[dvipsnames]{xcolor}
\graphicspath{{figures_v2/}}

\DeclareCaptionFont{xsmall}{\fontsize{8.5}{10}\selectfont}
\renewcommand{\vec}[1]{{\mathbf{#1}}}
\renewcommand{\a}{\alpha}
\renewcommand{\b}{\beta}
\newcommand{\ham}{\hat{\mathcal{H}}}
\newcommand{\gm}{\vec{G}_\mathrm{m}}

\begin{document}

\preprint{APS/123-QED}

\title{Relaxation-driven flat bands and topology in moir\'e transition metal dichalcogenide heterobilayers}

\author{Mitchell Luskin}
\affiliation{School of Mathematics, University of Minnesota, Minneapolis, MN 55455, USA}

\author{Max Geier}
\affiliation{Department of Physics, Massachusetts Institute of Technology, Cambridge, MA 02139, USA}

\author{Liang Fu}
\affiliation{Department of Physics, Massachusetts Institute of Technology, Cambridge, MA 02139, USA}

\author{Ziyan Zhu}
\email[Corresponding author: ]{zhuziyan@bc.edu}
\affiliation{Department of Physics, Boston College, Chestnut Hill, MA 02472, USA}

\begin{abstract}
Moiré transition metal dichalcogenide (TMD) heterobilayers are commonly modeled by a continuum theory that yields topologically trivial bands, in contrast to their homobilayer counterparts which host topological bands and fractional Chern insulators (FCI). We show this conclusion is an artifact of neglecting the pseudomagnetic field generated by lattice relaxation, an effect intrinsic to every moir\'e material. We develop a continuum model that resolves relaxation into three channels: a modified moir\'e potential with higher Fourier harmonics, a pseudoelectric (scalar deformation) potential, and a pseudomagnetic (vector) potential. Using WSe$_2$/WS$_2$ as a prototype, we find that the pseudomagnetic field alone gaps the third and fourth valence bands with Chern numbers $\pm 1$ over a broad range of twist angle and lattice mismatch, while the moir\'e potential correction and pseudoelectric potential narrow the bandwidth and enhance the bandgaps, which survive many-body interactions using neural-network variational Monte Carlo calculations. Relaxation also smoothens the Berry curvature and quantum metric relative to the rigid model, moving the band closer to the ideal Chern limit, beneficial for the quantum anomalous Hall effect, FCI states, and flat-band superconductivity when filled to higher bands. Our work establishes a new framework that connects first-principles calculations, through the continuum model, to many-body observables. Using this framework, we show moir\'e heterobilayers as a new class of topological materials whose topology is driven entirely by intrinsic lattice relaxation. 
\end{abstract}

\maketitle


\emph{Introduction ---} Lattice relaxation, the spontaneous atomic reconstruction that minimizes the sum of intralayer elastic energy and the interlayer misfit energies, is a defining feature of moir\'e superlattices~\cite{dai2016twisted, nam2017lattice, carr2018relaxation, naik2018ultra}. At small twist angles or lattice mismatches, relaxation drives the formation of large, uniform triangular or hexagonal domains separated by sharp domain walls, fundamentally reshaping the electronic landscape~\cite{nam2017lattice, naik2018ultra, yoo2019atomic, carr2020exact, enaldiev2020stacking, leconte2022relaxation, li2021imaging, vanwinkle2023rotational, klein2026imaging, weston2020atomic}. In transition metal dichalcogenide (TMD) heterobilayers such as WSe$_2$/WS$_2$, this reconstruction underlies the observation of Mott insulators~\cite{tang2020simulation, regan2020mott, huang2020correlated}, generalized Wigner crystals~\cite{regan2020mott, huang2020correlated}, and moir\'e excitons~\cite{jin2019observation, tran2019evidence}. 

Despite reshaping electronic structures, relaxation is treated incompletely in all existing continuum models for these systems~\cite{wu2018hubbard, wu2019topological, angeli2021gamma, devakul2021magic, zhang2024polarization, mao2024transfer, zhang2025twist}. 
When relaxation is accounted for in the existing continuum model~\cite{angeli2022twistronics}, relaxed moir\'e potential is considered, but the strain-induced terms are neglected.
The spatially varying strain $\epsilon_{ij}(\vec{r})$ produces a scalar deformation potential (pseudoelectric field) and a vector potential (pseudomagnetic field) that have no counterpart in the rigid model. 
These strain-induced pseudo-gauge field can have physical consequences:  
relaxation-driven pseudomagnetic field reshapes flat bands in twisted bilayer graphene~\cite{nam2017lattice,carr2020exact}
and periodic strain fields in heterobilayer TMDs and heterostrain in twisted bilayer graphene give rise to nontrivial topology~\cite{xie2022valley,bi2019designing}. 
However, whether this mechanism survives in realistic heterobilayers and how it competes with other relaxation-induced effects has remained open. To our knowledge, no continuum model for moir\'e TMD heterobilayers has incorporated the relaxation effects with the full set of strain-induced gauge fields.

Nontrivial band topology in moir\'e TMDs has been predicted and observed in homobilayers~\cite{wu2019topological,devakul2021magic,reddy2023fractional,crepel2023anomalous,
wang2024fractional, foutty2024mapping,zeng2023thermodynamic,park2023observation,xu2023observation, cai2023signatures}. 
In heterobilayers, MoTe$_2$/WSe$_2$ exhibits a quantum anomalous Hall effect at half filling of the top band ($\nu=1$) at finite displacement fields due to intertwined bands localized on different layers~\cite{li2021quantum}.
Topological phases at zero displacement field, at higher band fillings, or with the near-ideal quantum geometry required to stabilize FCI states have not been reported in any realistic heterobilayer model. 

\begin{figure*}[ht!]
    \centering
    \includegraphics[width=\linewidth]{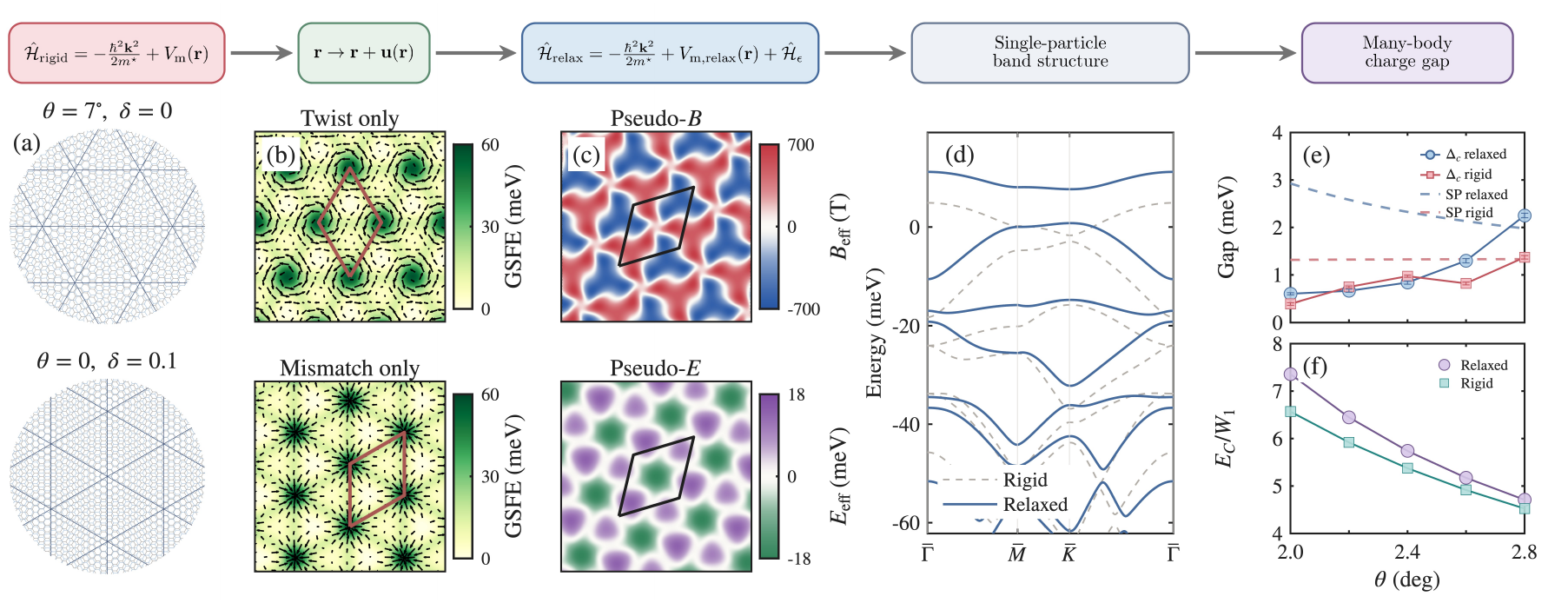}
    \caption{Overview of the relaxed continuum model. (a) Moir\'e patterns formed by pure twisting ($\theta = 7^\circ$, top) and by pure lattice mismatch ($\delta = 0.1$, bottom). (b) Relaxation pattern for twist only ($\theta = 2^\circ$, $\delta = 0$, top) and for lattice mismatch only ($\theta = 0^\circ$, $\delta = 0.02$, bottom). Color indicates the generalized stacking fault energy (GSFE); arrows show the relative relaxation displacement $\vec{u}_1 - \vec{u}_2$; parallelograms mark the moir\'e unit cell. (c) Relaxation-induced pseudomagnetic field $B_\mathrm{eff}$ (top) and pseudoelectric field $E_\mathrm{eff}$ (bottom), both evaluated at $\theta = 1.2^\circ$, $\delta = 0.02$. (d) Relaxed (solid) and rigid (dashed) band structures with the same parameters as (c). (e) Many-body charge gap $\Delta_c = E(N{+}1) + E(N{-}1) - 2E(N)$ at filling $\nu = 2$, i.e., $N = 18$ particles in a triangular supercell of $3 \times 3$ unit cells, at fixed lattice mismatch $\delta = 0.02$, from neural-network variational Monte Carlo~\cite{SM} for the relaxed (blue circles) and rigid (red squares) models; error bars denote the Monte Carlo standard error and are comparable to the symbol size. Dashed lines show the corresponding single-particle band gaps of the two models. (f) Ratio of the Coulomb energy $E_C = e^2/4\pi\epsilon\epsilon_0 r_0(\theta)$ at the mean interparticle distance $r_0$ of the $\nu = 2$ cell (with $\epsilon = 5$) to the bandwidth $W_1$ of the topmost band, for the same two models. The ratio decreases monotonically with twist angle, so interaction effects are strongest at small $\theta$, and is larger in the relaxed model at every angle.}
    \label{fig:schematic}
\end{figure*}

In this Letter, we develop a continuum model for TMD heterobilayers that incorporates all three channels through which lattice relaxation enters the Hamiltonian: the modified moir\'e potential, a pseudoelectric field, and a pseudomagnetic field. 
Using WSe$_2$/WS$_2$ as a model system, we show that (1) the pseudoelectric potential and the higher-order harmonics of the moir\'e potential flattens the top two bands and increase their bandgap; (2) the pseudomagnetic field opens a topological gap between the third and fourth valence bands, with Chern numbers $C_3 = +1$ and $C_4 = -1$, predicting nontrivial topological states at $\nu = 6$; and (3) relaxation smoothens the Berry curvature and quantum metric toward the ideal Chern limit, beneficial for FCI states. The model forms a framework that connects first-principles calculations to many-body observables [Fig.~\ref{fig:schematic}]: density functional theory (DFT)-parametrized stacking energetics determine the relaxed geometry, the relaxed geometry fixes the continuum Hamiltonian, and the resulting bands serve as direct input to many-body methods, which we demonstrate by computing the relaxation-enhanced charge gap with neural-network variational Monte Carlo \cite{carleo2017,Schutt2017Schnet,Luo2019backflow,Pfau2020Sep,hermann2020,choo2020,vonGlehn2022Nov,Viteritti2023,gao2023generalizing,casella2023,wilson2023,smith2024,RLi2024forwardlaplacian,Pescia2024,kim2024,Geier2025Attn,avdoshkin2025spin,foster2025orbformer}. The framework is generalizable to arbitrary heterobilayer TMD systems.


\emph{Model} --- A twist angle $\theta$ and/or lattice mismatch $\delta$ between two layers produces a moir\'e superlattice [Fig.~\ref{fig:schematic}(a)]. In heterobilayer TMDs, both twist angle and lattice constant mismatch can coexist and generate different relaxation patterns [Fig.~\ref{fig:schematic}(b)]: twist alone produces a rotational displacement field, whereas lattice mismatch alone produces a divergent pattern. In both cases, relaxation expands the equilibrium stacking domains by either locally twisting or stretching (see Supplemental Material for details on the continuum relaxation model~\cite{SM}).

In TMD heterobilayers such as WSe$_2$/WS$_2$, the valence band maximum at the $K$-point is dominated by the higher-lying layer, and in this case, WSe$_2$. The Hamiltonian can therefore be written as a WSe$_2$ low-energy Hamiltonian in a moir\'e potential and a momentum-dependent strain term: 
\begin{align}
\hat{\mathcal{H}} = \hat{\mathcal{H}}_\mathrm{mono} + \hat{\mathcal{H}}_\mathrm{m} + \hat{\mathcal{H}}_\epsilon, \label{eqn:full_ham}
\end{align}
where 
\begin{equation}
\hat{\mathcal{H}}_\mathrm{mono} (\vec{q}) = -\frac{\hbar^2 \mathbf{q}^2}{2m^\star} c^\dagger_\mathbf{q} c_\mathbf{q},
\end{equation}
is the WSe$_2$ low-energy Hamiltonian with $m^\star = 0.55 m_e$ is the effective mass of WSe$_2$~\cite{fang2018electronic}. 

The second term is the moir\'e potential:
\begin{align}
\hat{\mathcal{H}}_\mathrm{m}(\vec{q}) = \sum_{\vec{G}_\mathrm{m}} V_\mathrm{m} (\vec{G}_\mathrm{m}) c^\dagger_{\vec{q} + \vec{G}_\mathrm{m}} c_\vec{q},
\end{align}
with $V_\mathrm{m}(-\vec{G}_\mathrm{m})=V_\mathrm{m}(\vec{G}_\mathrm{m})^\star$, which we parameterize with DFT calculations~\cite{wu2018hubbard,angeli2021gamma,angeli2022twistronics}. In the absence of relaxation, the moir\'e potential can be obtained from the valence band edge as a function of the relative shift between two layers, $\vec{b}$, which can be fitted to the functional form of $V_\mathrm{m}(\vec{b}) = \sum_{s,j} 2V_s \cos(\vec{G}_j^s \cdot \vec{b} + \phi_s)$, where $s$ is the shell index with  corresponding monolayer reciprocal lattice vectors $\vec{G}_j^s,$ and $V_s$ and $\phi_s$ are fitting parameters. There is a one-to-one mapping between the rigid shift (local configuration) and real space position in a moir\'e cell, $\vec{r}$~\cite{carr2018relaxation}. Relaxation introduces a relative displacement between two layers, $\Delta \vec{u}$, which modifies the local stacking to be $\vec{b} \rightarrow \vec{b} + \Delta \vec{u} (\vec{b})$ and $V_\mathrm{m, relax}(\vec{b}) = \sum_{s,j} 2V_s \cos(\vec{G}_j^s \cdot (\vec{b} + \Delta \vec{u}) + \phi_s)$. As a result, the moir\'e potential sharpens and forms domain walls, which require higher-shell expansions. Finally, we Fourier expand $V_\mathrm{m,relax} (\vec{b})$ up to the 15-th shell to obtain $V_\mathrm{m}(\vec{G}_\mathrm{m})$. 

The last term is the strain Hamiltonian: 
\begin{align}
    \ham_\epsilon (\vec{q}) = \sum_{\gm} \left[ \mathcal{C}\,\epsilon_0(\gm)
    - \mathcal{D}\,\vec{\epsilon}(\gm)\cdot\left(\vec{q} + \tfrac{1}{2}\gm\right) \right] \nonumber \\
    \times\, c_{\vec{q}+\gm}^\dagger c_{\vec{q}},
\end{align}
where $\epsilon_0 = \epsilon_{xx}+\epsilon_{yy}$ and $\vec{\epsilon} = (\epsilon_1,\epsilon_2)$, with $\epsilon_1 = \epsilon_{xx}-\epsilon_{yy}$ and $\epsilon_2 = -2\epsilon_{xy}$, are the strain components; $\mathcal{C} = -2.25$~eV and $\mathcal{D} = 8.32 \ \mathrm{eV\cdot \text{\AA}}$ are the strain coefficients for monolayer WSe$_2$, obtained from \citet{fang2018electronic} and details can be found in Supplemental Material~\cite{SM}. 

Defining a real-space moir\'e-scale envelope function 
$
\Psi(\vec{r}) = \frac{1}{\sqrt{|\Gamma^*_\mathrm{m}|}} \int_{\vec{q}} d^2\vec{q}\,e^{i\vec{q}\cdot\vec{r}}\, c_{\vec{q}},
$
where $\Gamma^*_\mathrm{m}$ is the moir\'e Brillouin zone area, we can rewrite $\hat{\mathcal{H}}_\mathrm{\epsilon}$ in the envelope function basis in the pseudo-gauge form: 
\begin{equation}
\ham_\epsilon = \int d^2\vec{r}\,\Psi^\dagger(\vec{r}) \left[\Phi(\vec{r}) +  \{\hat k_\a, A_\a(\vec{r})\}\right]\Psi(\vec{r}), 
\label{eqn:strain_gauge}
\end{equation}
where $\Phi(\vec{r}) = \mathcal{C} \epsilon_0 (\vec{r}) $ is the scalar deformation potential that modifies the moir\'e potential, $\vec{A}(\vec{r}) = -\frac{1}{2}\mathcal{D}\, (\epsilon_1, \epsilon_2)$ is the pseudo-gauge field that gives rise to a pseudomagnetic field $\vec{B} (\vec{r})= \nabla \times \vec{A}(\vec{r})$, and $\hat k_\a = -i \partial_\a$.

We mainly consider 3R stacking (parallel or $0^\circ$ twist).
Results for 2H stacking (antiparallel or $180^\circ$ twist) are presented in the Supplemental Material~\cite{SM}. 
The natural lattice mismatch between WSe$_2$ and WS$_2$ is $\delta \approx 0.04$. In this work, we treat $\delta$ as a free parameter. 

\begin{figure}[ht!]
    \centering
    \includegraphics[width=0.9\linewidth]{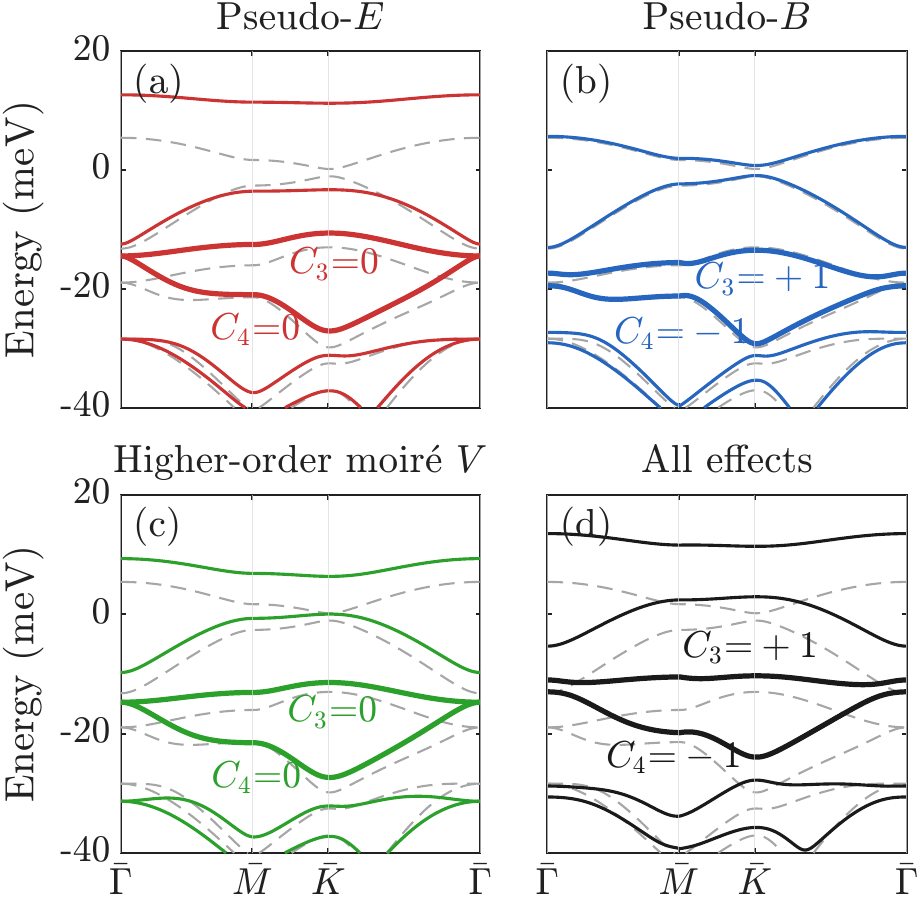}
    \caption{Decomposition of relaxation effects on the moir\'e band structure of WSe$_2$/WS$_2$ at $\theta = 1.0^\circ$, $\delta = 0.02$. Each panel shows the top valence bands (solid, colored) with the rigid band structure overlaid (gray dashed). (a) pseudoelectric field only, $\Phi \propto \epsilon_0$; (b) pseudomagnetic field only, $\vec{A} \propto (\epsilon_1, \epsilon_2)$; (c) relaxed moir\'e potential $V_\mathrm{m,\,relax}$ without strain; (d) all three channels combined. The third and fourth valence bands are drawn with increased linewidth, and their Chern numbers $C_3$, $C_4$ are labeled.}
    \label{fig:bands}
\end{figure}

\emph{Decomposition of relaxation effects} --- To identify which channel drives each feature of the relaxed band structure, we isolate the contribution of each term in Eq.~\eqref{eqn:full_ham}. Figure~\ref{fig:bands} shows the moir\'e band structure of 3R-stacked WSe$_2$/WS$_2$ at $\theta = 1.0^\circ$, $\delta = 0.02$ under four conditions: pseudoelectric field $\Phi$ only [Fig.~\ref{fig:bands}(a)], pseudomagnetic field $\vec{A}$ only [Fig.~\ref{fig:bands}(b)], the relaxed moir\'e potential $V_\mathrm{m,\,relax}$ with no strain [Fig.~\ref{fig:bands}(c)], and all three channels combined [Fig.~\ref{fig:bands}(d)].

The effect of the scalar deformation potential and higher-order moir\'e potential are similar: they both open up a band gap between the top two valence bands and reduce their bandwidths by localizing electrons at the AA spots, where the moir\'e potential has global maxima. Both corrections are momentum-independent and all gaps and degeneracies are preserved. In contrast, the pseudomagnetic field opens up a topological gap $\Delta_{34}$ between the third and fourth valence bands, with Chern numbers $C_3 = +1$ and $C_4 = -1$ [Fig.~\ref{fig:bands}(b)], which is consistent with the Chern number $C = \pm 1$ induced by topological band gap opening at the quadratic band touching points~\cite{chong2008effective,sun2009topological}.
The fully relaxed Hamiltonian inherits the flattened bands and increased band gaps, as well as the induced nontrivial topology [Fig.~\ref{fig:bands}(d)].

\begin{figure}[ht!]
    \centering
    \includegraphics[width=\linewidth]{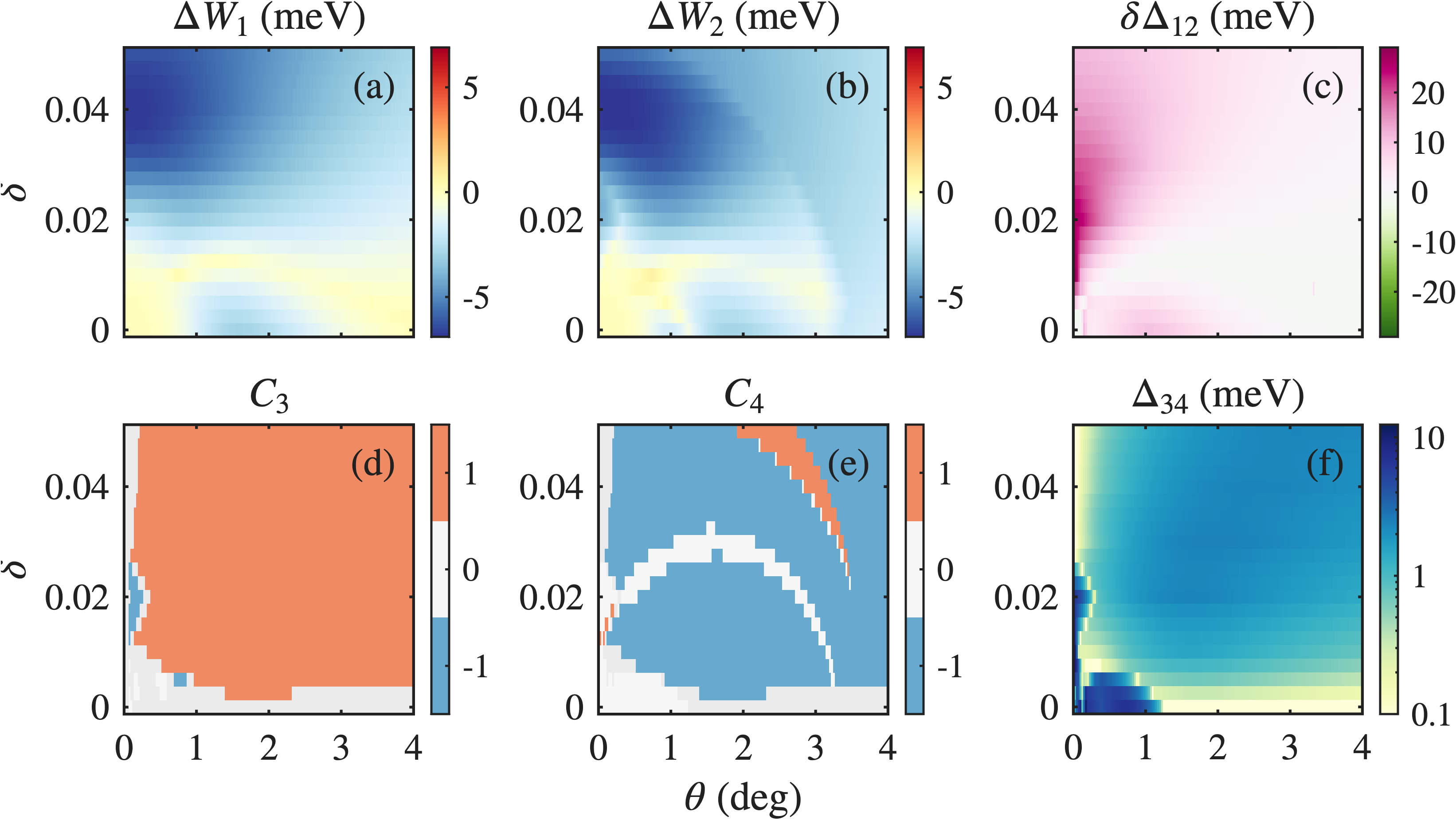}
    \caption{Relaxation-induced band structure changes and topology across the $(\theta, \delta)$ plane for 3R-stacked WSe$_2$/WS$_2$. Top row: changes of the electronic structures from the rigid to the relaxed model. (a) $\Delta W_1 = W_1^\mathrm{relax} - W_1^\mathrm{rigid}$ and (b) $\Delta W_2$ show bandwidth narrowing of the top valence bands. (c) $\delta\Delta_{12} = \Delta_{12}^\mathrm{relax} - \Delta_{12}^\mathrm{rigid}$ shows enhancement of the gap between the first and second valence bands. Bottom row: topology of the relaxed model. (d) Chern number $C_3$ of the third valence band, $C_3 = +1$ over the shaded nontrivial region. (e) Chern number $C_4$ of the fourth valence band, predominantly $C_4 = -1$. Gray regions: Chern number ill-defined ($\Delta_{34} < 0.3$~meV). (f) Gap $\Delta_{34}$ between the third and fourth valence bands on a logarithmic scale.}
    \label{fig:topo}
\end{figure}

\emph{Phase diagram} --- Relaxation-induced flat bands and topology are general across a wide range of lattice constant mismatches and twist angles [Fig.~\ref{fig:topo}]. The bandwidths of the top two valence bands, $W_1$ and $W_2$, are reduced relative to the rigid model over essentially the entire parameter range [Fig.~\ref{fig:topo}(a)-(b)], with the largest narrowing at small twist angles and modest lattice mismatch. At the same time, the gap $\Delta_{12}$ between the first and second valence bands is enhanced by relaxation [Fig.~\ref{fig:topo}(c)], further isolating the topmost band and supporting the single-band Hubbard model construction~\cite{wu2018hubbard,morales-duran2022nolocal}.

Figure~\ref{fig:topo}(d)-(e) maps the topology of the third and fourth valence bands across the $(\theta, \delta)$ plane for 3R-stacked WSe$_2$/WS$_2$, with gray regions indicating bands with adjacent bands within a 3~meV tolerance. 
The third valence band acquires a Chern number $C_3 = +1$ over a broad, connected region, bounded by gap closings. Within this region, a topological gap is opened by a pseudomagnetic field [Fig.~\ref{fig:topo}(f)]. 

The gap closes along a dome near $\delta = 0.02$ at small twist angles, separating two topologically distinct regions. Inside the dome, the second and third valence bands carry $C_2 = +1$ and $C_3 = -1$, respectively (see Supplemental Material~\cite{SM} for the topology of the second band). Outside of the dome, the second band becomes topologically trivial, $C_3$ changes sign to +1, and $C_4 = -1$ over a largely overlapping region. 
The nontrivial topology persists across the realistic lattice constant mismatch of WSe$_2$/WS$_2$ with $\delta \approx 0.04$, suggesting that quantum anomalous Hall states are experimentally accessible at $\nu=6$. 
The top two bands remain topologically trivial over the entire parameter range. 

The relaxation-induced gap enhancement survives beyond the single-particle
picture and directly accessible
to transport and compressibility measurements. Its many-body counterpart is the gauge-invariant charge gap
$\Delta_c = E(N{+}1) + E(N{-}1) - 2E(N)$, where $E(N)$ is the $N$-particle
ground-state energy from neural-network variational Monte Carlo.
Figure~\ref{fig:schematic}(e) compares $\Delta_c$ for the rigid and relaxed
models at $\nu = 2$ ($N = 18$ particles in a $3\times3$ supercell,
$\delta = 0.02$).
For $\theta \gtrsim 2.6^\circ$ the relaxed model has a systematically larger
gap, consistent with the single-particle results [Fig.~\ref{fig:topo}(c)];
by $\theta \sim 2^\circ$ the two become comparable. This crossover tracks the
interaction-to-bandwidth ratio [Fig.~\ref{fig:schematic}(f)]: with
$e^2/4\pi\epsilon\epsilon_0 r_0 \approx 100$~meV against
$W_1 = 10-20$~meV, interactions increasingly dominate at small twist
and wash out the distinction between rigid and relaxed models.


\emph{Conditions for FCI} --- In addition to a topological gap, an FCI state at partial filling of the third band requires the band's quantum geometry to resemble that of a Landau level~\cite{roy2014band,jackson2015geometric,wang2021exact,ledwidth2023vortexability}. 

The quantum geometry of a Bloch band is encoded in the quantum geometric tensor $\mathcal{G}_{\a\b} = \langle \partial_{k_\a} u | (\mathbb{I} - |u\rangle\langle u|) | \partial_{k_\b} u \rangle$, whose imaginary part gives the Berry curvature $\Omega = -2\,\mathrm{Im}\,\mathcal{G}_{xy}$ and whose real part gives the quantum metric $g_{\a\b} = \mathrm{Re}\,\mathcal{G}_{\a\b}$. 
In addition to a small bandwidth, closeness to the ideal limit is captured by two diagnostics~\cite{morales-duran2023pressure,dong2024anomalous,wang2021exact,xu2024maximally,ledwidth2024fractional}: the integrated trace-condition violation 
\begin{equation}
    \frac{T}{2\pi} \equiv \frac{1}{2\pi}\!\int (\mathrm{Tr}\,g - |\Omega|)\, d^2q \geq 0, 
\end{equation}
which vanishes for an ideal band, and the Berry curvature fluctuation
\begin{equation}
    F = \sqrt{\langle (\tilde\Omega - C)^2\rangle_\mathrm{BZ}}, 
\end{equation}
where $\tilde\Omega = \Omega \frac{A_\mathrm{BZ}}{2\pi}$ is the dimensionless per-moir\'e cell Berry curvature. 
Smaller $T/2\pi$ and $F$ indicate stronger FCI stability. 
As a benchmark, the first moir\'e valence band of twisted bilayer MoTe$_2$, where FCI states have been experimentally observed, has $T/2\pi$ and $F$ of order unity~\cite{xu2024maximally}.

\begin{figure}[ht!]
    \centering
    \includegraphics[width=\linewidth]{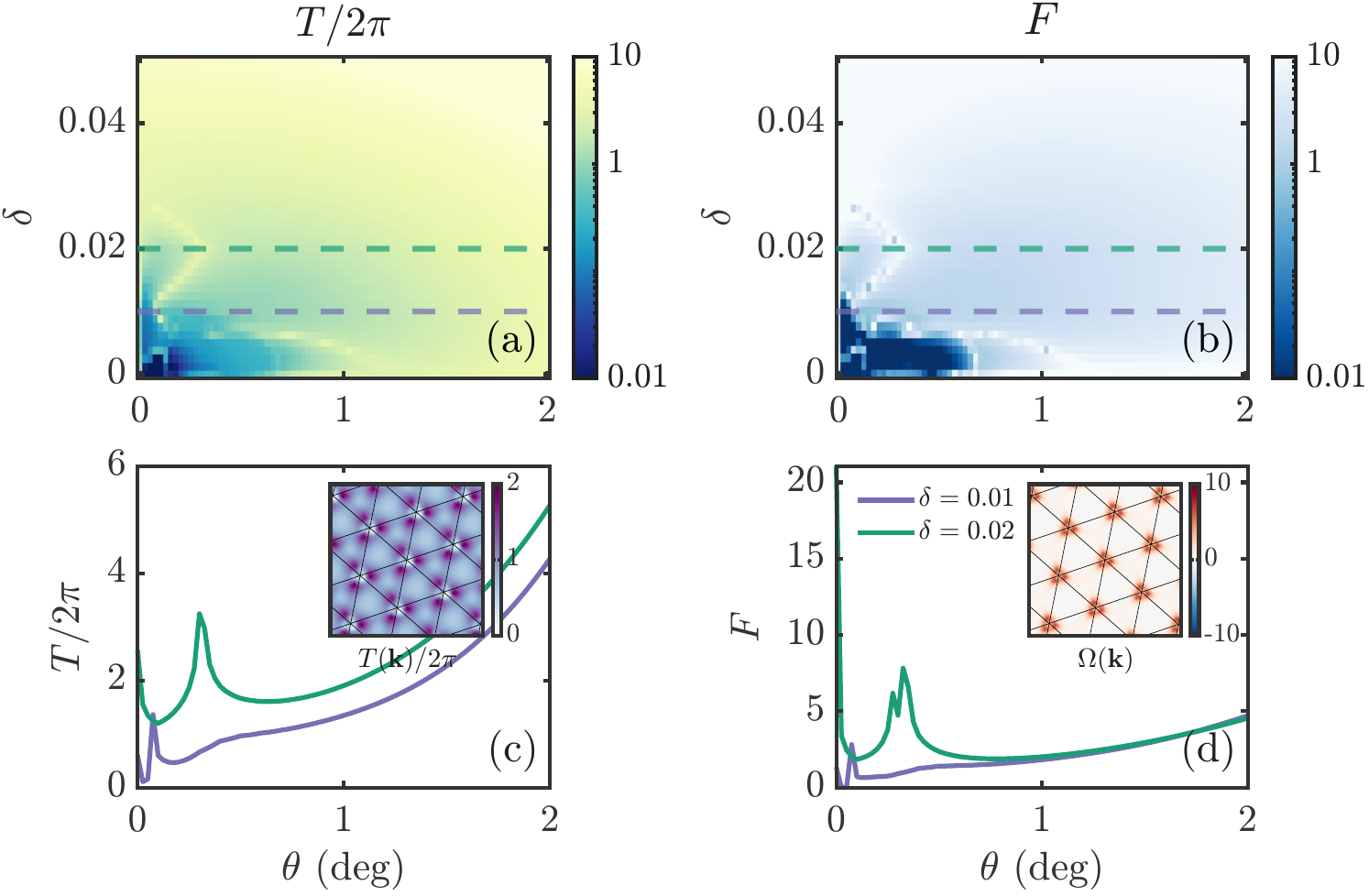}
    \caption{Ideal band condition of the third valence band across the 3R-stacked relaxed model. (a) Integrated trace-condition violation $T/2\pi = (2\pi)^{-1}\!\int(\mathrm{Tr}\,g - |\Omega|)\,d^2q$ as a function of lattice constant mismatch and twist angle. (b) Berry-curvature fluctuation $F = \sqrt{\langle (\tilde\Omega - C)^2\rangle_\mathrm{BZ}}$, with $\tilde\Omega = \Omega\,A_\mathrm{BZ}/(2\pi)$. Dashed horizontal lines mark the line cuts shown below. (c) $T/2\pi$ versus twist angle at $\delta = 0.01$ (green) and $\delta = 0.02$ (purple). Inset: pointwise $T(\vec{q})/2\pi$ over the moir\'e Brillouin zone at $\theta = 0.5^\circ$, $\delta = 0.01$. (d) Same as (c) for $F$, with inset showing the dimensionless Berry curvature $\Omega(\vec{q}) \equiv \tilde\Omega(\vec{q})$ at $\theta = 0.5^\circ$, $\delta = 0.01$. Black lines outline the moir\'e Brillouin zone tiling.}
    \label{fig:geom}
\end{figure}

Figure~\ref{fig:geom}(a)-(b) maps both diagnostics across the $(\theta,\delta)$ plane for the relaxed third valence band. Both are minimized at small twist angles and modest mismatch ($\delta \sim 0.01-0.02$), and rise monotonically outside of it. 
The topological phase boundary identified in Fig.~\ref{fig:topo}(d)-(e) is also visible as a ridge of enhanced $T/2\pi$ and $F$. Both diagnostics are small on either side of the gap closing, making this parameter range a viable FCI candidate. 
Line cuts at $\delta = 0.01$ and $\delta = 0.02$ [Fig.~\ref{fig:geom}(c)-(d)] show that both quantities pass through two local minima, separated by the gap closing peaks at $\theta \sim 0.1^\circ$ and $\theta \sim 0.3^\circ$ respectively. 
At $\delta = 0.01$, both diagnostics are less than 1 for $\theta$ between $0.1^\circ$ and $0.5^\circ$, an order of magnitude smaller than the rigid model. 

The momentum-space maps of $T(\vec{q})/2\pi$ and $\Omega(\vec{q})$ at a representative point ($\theta = 0.5^\circ$, $\delta = 0.01$) [insets of Fig.~\ref{fig:geom}(c)-(d)] show that the point-wise trace violation condition is relatively uniform across momenta, while the Berry curvature fluctuation, though still concentrated at the Brillouin zone corner, is substantially smoothened relative to the rigid limit (see Supplemental Material~\cite{SM}).

These results identify small-mismatch TMD heterobilayers as the most promising platform for FCI states. At the natural mismatch of WSe$_2$/WS$_2$ ($\delta \sim 0.04$), the band remains topological, though $T/2\pi$ is larger than in twisted MoTe$_2$. Pairs with smaller lattice constant mismatch such as WSe$_2$/MoSe$_2$ ($\delta\sim0.003$) and WSe$_2$/MoTe$_2$ ($\delta\sim0.007$), fall within the most favorable region of Fig.~\ref{fig:geom}(a)-(b) and may approach the ideal Chern limit at experimentally accessible twist angles. 


\emph{Discussion} --- We have developed a continuum model for moir\'e TMD heterobilayers that fully incorporates lattice relaxation. We identify the pseudomagnetic gauge field as the sole driver of band topology: it opens a gap between the third and fourth valence bands, while the pseudoelectric and higher-harmonic moir\'e channels narrow the bandwidth and tune the inter-band gap structure. This decomposition explains why previous continuum models, which either neglected strain-induced gauge fields or bundled relaxation into a single effective potential, found trivial topology.

The physics here is distinct from both graphene, where pseudomagnetic fields require externally applied strain~\cite{guinea2010energy, levy2010strain, low2010strain, vozmediano2010gauge}, and TMD homobilayers, where topology arises already in the rigid limit~\cite{devakul2021magic, wang2024fractional}. In heterobilayers, the strain-induced gauge fields are an unavoidable consequence of atomic reconstruction, making the resulting topology intrinsic.

Our framework is generalizable to any heterobilayer TMD system: the model structure is dictated by symmetry and only the DFT-derived moir\'e potential and tight-binding parameters are material-dependent. Since different TMD pairs span a range of lattice mismatches $\delta$, the phase diagrams presented here provide a guide for identifying candidate systems with optimal ideal topological bands. 

Our results make several experimentally testable predictions. First, at integer filling $\nu = 6$ of the moir\'e valence bands, a quantum anomalous Hall state should emerge with quantized Hall conductance $\sigma_{xy} = e^2/h$. Existing experiments observe a trivial correlated insulator at $\nu = 1$ in WSe$_2$/WS$_2$~\cite{tang2020simulation, regan2020mott}, consistent with our finding that the topmost valence band is trivial. The relaxation enhancement of the $\nu = 2$ charge gap [Fig.~\ref{fig:schematic}(e)] provides a further quantitative test. Second, at fractional fillings of the third valence band, the combination of flat bandwidth, nontrivial Chern number, and favorable quantum geometry suggests the possibility of FCI states.

\emph{Acknowledgments} --- We thank Ying Ran, Qirui Ren, and Ben Feldman for helpful discussion. 
ZZ's work is supported by a startup fund from Boston College. 
ML's research was partially supported by Simons Targeted Grant Award No. 896630. ML's research was also supported in part by NSF Award No. 2422469. 
MG and LF were supported by a Simons Investigator Award from the Simons Foundation.
A portion of the research used computational resources at the Boston College Andromeda High Performance Computing cluster. The research also used resources of the National Energy Research Scientific Computing Center (NERSC), a Department of Energy User Facility, using NERSC award ERCAP0039049.

{\em Code availability.---}
The NNVMC calculations in this work are based on our public codebase {\tt PeriodicWave} \cite{periodicwave_github}, described in Ref.~\cite{Geier2025Attn}.

{\em Data availability.---}
The data required to reproduce and benchmark the NNVMC results, specifically the ground state energies of the wavefunctions used to generate the numerical results in Fig.~\ref{fig:schematic}(e), are contained in the Supplemental Material. Code and data will be available upon reasonable request. 

\bibliography{ref}

\end{document}


\title{Supplemental Material for ``Relaxation-driven band topology in moir\'e transition metal dichalcogenide heterobilayers''}

\maketitle

\setcounter{equation}{0}
\setcounter{figure}{0}
\setcounter{table}{0}
\setcounter{section}{0}
\renewcommand{\theequation}{S\arabic{equation}}
\renewcommand{\thefigure}{S\arabic{figure}}
\renewcommand{\thetable}{S\arabic{table}}
\renewcommand{\thesection}{S\arabic{section}}
\renewcommand{\thesubsection}{S\arabic{section}.\arabic{subsection}}
\renewcommand{\thesubsubsection}{S\arabic{section}.\arabic{subsection}.\arabic{subsubsection}}

\tableofcontents

\vspace{1em}
\noindent This Supplemental Material is organized as follows. Section~\ref{sec:derivation} derives the relaxed continuum model Hamiltonian from a tight-binding starting point. Section~\ref{sec:param} presents the model parametrization, including the Schrieffer-Wolff projection onto the valence band and the DFT-based moir\'e potential with its rigid and relaxed Fourier decompositions. Section~\ref{sec:relaxation} describes the continuum lattice relaxation model, including the generalized stacking fault energy (GSFE), the configuration-space formulation, and the energy minimization procedure. Section~\ref{sec:results} presents supplemental results including emergent gauge fields, band structures at multiple twist angles, bandwidth and bandgap phase diagrams, superfluid stiffness, and corresponding results for 2H (180$^\circ$) stacking.


\section{Continuum model derivation and computational details}
\label{sec:derivation}

\subsection{Tight-binding Hamiltonian and Fourier transform}

We start from the real-space tight-binding Hamiltonian
\begin{equation}
\ham = \sum_{\vec{R}, \Delta\vec{r}, s,s'} t^{ss'}(\Delta\vec{r})\, c_{\vec{R}+\Delta\vec{r}}^\dagger c_{\vec{R}},
\end{equation}
where $\vec{R}$ labels Bravais lattice sites, $\Delta\vec{r}$ is the bond vector summing over all near neighbors (both $\pm\Delta\vec{r}$) and Hermiticity is guaranteed by $t(-\Delta \vec{r}) = t(\Delta \vec{r})^*$, and $s, s'$ are sublattice indices. We decompose the hopping into a local component and a long-wavelength moir\'e modulation:
\begin{equation}
t^{ss'}(\Delta\vec{r}; \vec{R}) = t_0^{ss'}(\Delta\vec{r}) + \sum_{\gm} e^{i\gm \cdot (\vec{R} + \Delta\vec{r}/2)} t(\gm),
\end{equation}
where $\gm$ is a moir\'e reciprocal lattice vector. Note that $\exp(i\gm \cdot \vec{R}) \neq 1$ in general, since $\gm$ is not a reciprocal lattice vector of the monolayer. Throughout, we associate slowly varying moir\'e-scale fields (the modulation here and the strain field below) with the bond midpoint $\vec{R} + \Delta\vec{r}/2$ --- the symmetric choice shared by the two hopping directions --- which makes each term manifestly Hermitian.

We Fourier-transform the electron operators with the convention
\begin{equation}
c_{\vec{R}} = \frac{1}{\sqrt{|\Gamma^*|}} \int d^2\vec{k}\, e^{i\vec{k}\cdot\vec{R}}\, c_{\vec{k}}, \qquad
c_{\vec{k}} = \frac{1}{\sqrt{|\Gamma^*|}} \sum_{\vec{R}} e^{-i\vec{k}\cdot\vec{R}}\, c_{\vec{R}},
\end{equation}
where $|\Gamma^*|$ is the monolayer reciprocal lattice area. Substituting into the tight-binding Hamiltonian gives
\begin{align}
\ham &= \frac{1}{|\Gamma^*|} \sum_{\vec{k}, \vec{k}'} \sum_{\vec{R}, \Delta\vec{r}}
 e^{-i\vec{k}\cdot(\vec{R} + \Delta\vec{r})\,}\,e^{i\vec{k}'\cdot\vec{R}}\, t(\Delta\vec{r}; \vec{R})\, c_{\vec{k}}^\dagger c_{\vec{k}'},
\end{align}

After inserting the decomposition of $t(\Delta\vec{r}; \vec{R})$, the local piece carries an $\vec{R}$-phase $e^{i(\vec{k}-\vec{k}')\cdot\vec{R}}$, while the moir\'e-modulated piece carries $e^{i(\vec{k}-\vec{k}'+\gm)\cdot\vec{R}}$ together with the bond-midpoint phase $e^{i\gm\cdot\Delta\vec{r}/2}$, which combines with the form-factor phase to shift its momentum argument by $\tfrac{1}{2}\gm$. Using the Bravais-lattice sum rule
\begin{equation}
\sum_{\vec{R}} e^{i\vec{k}\cdot\vec{R}} = |\Gamma^*| \sum_{\vec{G}} \delta(\vec{k} - \vec{G}),
\label{eqn:sum_rule}
\end{equation}
where $\vec{G}$ runs over monolayer reciprocal lattice vectors, and restricting momenta to the first Brillouin zone ($\vec{G} = 0$), the sum over $\vec{R}$ collapses the double momentum sum: the local term is diagonal in $\vec{k}$, and the moir\'e term couples $\vec{k}$ to $\vec{k}+\gm$. We obtain
\begin{align}
\ham &= \sum_{\vec{k}} \mathcal{E}(\vec{k})\, c_{\vec{k}}^\dagger c_{\vec{k}} + \sum_{\vec{k},\gm} \sum_{\Delta\vec{r}} e^{i(\vec{k}+\gm/2)\cdot\Delta\vec{r}}\, t(\gm)\, c_{\vec{k}+\gm}^\dagger c_{\vec{k}},
\label{eqn:h_kspace}
\end{align}
where $\mathcal{E}(\vec{k}) = \sum_{\Delta\vec{r}} e^{i\vec{k}\cdot\Delta\vec{r}} t_0^{ss'}(\Delta\vec{r})$ is the monolayer dispersion. The first term is the local band structure, while the second is the moir\'e potential.

\subsection{Modification of hopping by relaxation}

With relaxation, the intralayer bond vector $\Delta\vec{r}$ connecting two atoms in the same layer (both displaced by $\vec{u}_\ell$) is modified to
\begin{equation}
\Delta\vec{r}' \approx \left[\mathbb{I} + \nabla\vec{u}(\vec{R} + \frac{\Delta\vec{r}}{2}) \right]\,\Delta\vec{r},
\end{equation}
where we used the smoothness of $\vec{u}$ on the atomic scale. 
Expanding the local hopping to first order in the strain:
\begin{equation}
t_0^{ss'}(\Delta\vec{r}') \approx t_0^{ss'}(\Delta\vec{r}) + (\Delta\vec{r}\cdot\nabla\vec{u}) \cdot \nabla_{\vec{r}} t_0^{ss'}(\Delta\vec{r}).
\end{equation}
Writing $r_i \equiv (\Delta\vec{r})_i$ for the components and $r \equiv |\Delta\vec{r}|$, the first-order correction is
\begin{equation}
\delta t_0^{ss'} = \sum_{ij} u_{ij}\, r_j\, \frac{\partial t_0^{ss'}}{\partial r_i},
\end{equation}
where $u_{ij} = \partial u_i / \partial r_j$ decomposes into a symmetric strain $\epsilon_{ij} = (u_{ij} + u_{ji})/2$ and an antisymmetric rotation $\omega_{ij} = (u_{ij} - u_{ji})/2$. Since $\omega_{ij} r_i r_j = 0$, only the strain tensor contributes. Under the two-center approximation $t_0^{ss'}(\Delta\vec{r}) = \mathcal{T}(r, \hat{n})$ and neglecting the angular correction, the hopping correction reduces to
\begin{equation}
\delta t_0^{ss'} = \epsilon_{ij} \left(\vec{R} + \frac{\Delta \vec{r}}{2} \right)\, \frac{r_i r_j}{r}\, \frac{\partial \mathcal{T}}{\partial r}.
\end{equation}

\subsection{Strain Hamiltonian in momentum space}

With the modified hopping, the strain correction to the Hamiltonian is
\begin{equation}
\ham_\epsilon = \sum_{\vec{R},\Delta\vec{r}} \epsilon_{ij}\left (\vec{R} + \frac{\Delta \vec{r}}{2} \right)\, \frac{r_i r_j}{r}\, \frac{\partial \mathcal{T}}{\partial r}\, c_{\vec{R}+\Delta\vec{r}}^\dagger c_{\vec{R}}.
\end{equation}
We associate the smooth strain field with the bond midpoint, which makes $\hat{\mathcal{H}}_\epsilon$ Hermitian. However, note that in the continuum relaxation model (see Sec.~\ref{sec:relaxation}), there is no individual atomic
positions and the strain enters only through Fourier coefficient $\epsilon(\vec{G}_\mathrm{m}).$
Since the relaxation is smooth on the moir\'e scale, the strain field can be expanded in moir\'e Fourier components:
\begin{equation}
\epsilon_{ij}(\vec{r}) = \sum_{\gm} e^{i\gm\cdot\vec{r}}\, \epsilon_{ij}(\gm).
\label{eqn:eij_ft}
\end{equation}
Substituting $\epsilon_{ij}(\vec{R}+\tfrac{\Delta\vec{r}}{2}) = \sum_{\gm} \epsilon_{ij}(\gm)\, e^{i\gm\cdot\vec{R}}\, e^{i\gm\cdot\Delta\vec{r}/2}$, Fourier-transforming, and using $\sum_{\vec{R}} e^{i(\vec{k}-\vec{k}'+\gm)\cdot\vec{R}} = |\Gamma^*|\,\delta(\vec{k}-\vec{k}'+\gm)$ to collapse the sum over $\vec{R}$, the midpoint phase $e^{i\gm\cdot\Delta\vec{r}/2}$ shifts the momentum argument of the form factor by $\tfrac{1}{2}\gm$, and we obtain
\begin{equation}
\ham_\epsilon = \sum_{\vec{k},\gm} F_{ij} \left(\vec{k} + \frac{1}{2} \vec{G}_\mathrm{m}\right)\, \epsilon_{ij}(\gm)\, c_{\vec{k}+\gm}^\dagger c_{\vec{k}},
\label{eqn:strain_ft}
\end{equation}
where we defined the strain form factor
\begin{equation}
F_{ij}(\vec{k}) \equiv \sum_{\Delta\vec{r}} e^{i\vec{k}\cdot\Delta\vec{r}}\, \frac{r_i r_j}{r}\, \frac{\partial \mathcal{T}}{\partial r}.
\end{equation}

\subsection{Low-energy expansion}

We expand all terms around the valence band maximum $\vec{k}_0$, writing $\vec{k} = \vec{k}_0 + \vec{q}$.

\subsubsection{Monolayer Hamiltonian}

For the monolayer dispersion, since $\Delta\mathbf{r}$ runs over $\pm$ pairs, $\mathcal{E}(\mathbf{k}) = \sum_{\Delta\mathbf{r}} e^{i\mathbf{k}\cdot\Delta\mathbf{r}}\, t_0$ is real; the $\pm$ pairs combine into the $2\cos$ form:
\begin{equation}
\mathcal{E}(\vec{k}) = \sum_{\Delta\vec{r}} 2\cos(\vec{k}\cdot\Delta\vec{r})\, t_0^{ss'}(\Delta\vec{r}),
\end{equation}
where the sum now runs over one member of each $\pm\Delta\vec{r}$ pair (three bonds for the first shell).
Taylor-expanding around $\vec{k}_0$:
\begin{equation}
\mathcal{E}(\vec{k}_0 + \vec{q}) = \mathcal{E}(\vec{k}_0) + q_i \left.\frac{\partial \mathcal{E}}{\partial k_i}\right|_{\vec{k}_0} + \frac{1}{2} q_i q_j \left.\frac{\partial^2 \mathcal{E}}{\partial k_i \partial k_j}\right|_{\vec{k}_0} + \cdots
\end{equation}
At the band extremum, the first derivative vanishes: $\partial \mathcal{E}/\partial k_i|_{\vec{k}_0} = 0$. The second-order term defines the effective mass tensor $m_{ij}^{-1} = -\hbar^{-2}\,\partial^2 \mathcal{E}/\partial k_i \partial k_j|_{\vec{k}_0}$, where the minus sign reflects the expansion around the valence band maximum (hole effective mass is positive). Under $C_3$ symmetry, the mixed term $q_x q_y$ is forbidden, and the effective mass is isotropic: $m_{xx} = m_{yy} \equiv m^*$. The monolayer Hamiltonian is therefore
\begin{equation}
\ham_\mathrm{mono} = -\frac{\hbar^2 (q_x^2 + q_y^2)}{2m^*}. \label{eqn:hlocal}
\end{equation}

\subsubsection{Strain Hamiltonian}

Expanding the strain form factor in Eq.~\eqref{eqn:strain_ft} around $\vec{k}_0$ to linear order in the momentum measured from $\vec{k}_0$, which is now $\vec{q}+\tfrac{1}{2}\gm$:
\begin{equation}
F_{ij}\left(\vec{k}_0 + \vec{q}+\frac{1}{2}\vec{G}_\mathrm{m} \right) \approx C_{ij} + \left(\vec{q}+\frac{1}{2}\vec{G}_\mathrm{m} \right)_\a D_{\a ij},
\end{equation}
where
\begin{align}
C_{ij} &= \sum_{\Delta\vec{r}} e^{i\vec{k}_0\cdot\Delta\vec{r}}\, \frac{r_i r_j}{r}\, \frac{\partial \mathcal{T}}{\partial r}, \qquad
D_{\a ij} = i \sum_{\Delta\vec{r}} e^{i\vec{k}_0\cdot\Delta\vec{r}}\, \frac{r_\a r_i r_j}{r}\, \frac{\partial \mathcal{T}}{\partial r}.
\end{align}
Over the full neighbor star, both $C_{ij}$ and $D_{\a ij}$ are real: the $C$ summand is even under $\Delta\vec{r} \to -\Delta\vec{r}$ while the $D$ summand is odd, so the $\pm$ pairs close into cosine and sine sums, respectively ($D_{\a ij}$ is real despite the explicit $i$, which multiplies a purely imaginary sum). This guarantees that the pseudogauge field defined below is real. The low-energy strain Hamiltonian is therefore
\begin{equation}
\ham_\epsilon = \sum_{\vec{q},\gm} \left [C_{ij} + \left(\vec{q} + \frac{1}{2} \vec{G}_\mathrm{m} \right)_\a D_{\a ij} \right]\, \epsilon_{ij}(\gm)\, c_{\vec{q}+\gm}^\dagger c_{\vec{q}}.
\label{eqn:strain_q}
\end{equation}

\subsubsection{Moir\'e potential}

For the moir\'e potential, the coupling in Eq.~\eqref{eqn:h_kspace} involves the form factor $F_m(\vec{k}) = \sum_{\Delta\vec{r}} e^{i\vec{k}\cdot\Delta\vec{r}}$ evaluated at the midpoint momentum $\vec{k}+\tfrac{1}{2}\gm$, where the sum runs over the $C_3$-symmetric star of monolayer neighbor vectors. Taylor-expanding around $\vec{k}_0$:
\begin{equation}
F_m\left(\vec{k}_0 + \vec{q} + \tfrac{1}{2}\gm\right) = F_m(\vec{k}_0) + \left(\vec{q} + \tfrac{1}{2}\gm\right)_i \left.\frac{\partial F_m}{\partial k_i}\right|_{\vec{k}_0} + \cdots
\end{equation}
The linear term vanishes by $C_3$ symmetry:
\begin{equation}
\left.\frac{\partial F_m}{\partial k_i}\right|_{\vec{k}_0} = i\sum_{\Delta\vec{r}} r_i\, e^{i\vec{k}_0\cdot\Delta\vec{r}} = 0,
\end{equation}
since $r_i \equiv (\Delta\vec{r})_i$ transforms as a vector under $C_3$ while the sum is over a $C_3$-invariant set. Retaining only the leading (constant) contribution and absorbing it into a new coefficient,
\begin{equation}
V_{\mathrm{m}}(\gm) \equiv F_m(\vec{k}_0)\, t(\gm) = \left[\sum_{\Delta\vec{r}} e^{i\vec{k}_0\cdot\Delta\vec{r}}\right] t(\gm),
\end{equation}
the low-energy moir\'e potential Hamiltonian reads
\begin{equation}
\ham_{\text{moir\'e}} = \sum_{\vec{q},\gm} V_{\mathrm{m}}(\gm)\, c_{\vec{q}+\gm}^\dagger c_{\vec{q}}. 
\label{eqn:hmoire}
\end{equation}
Here $V_{\mathrm{m}}(\gm)$ is the Fourier component of the real-space moir\'e potential, $V_{\mathrm{m}}(\vec{r}) = \sum_{\gm} V_{\mathrm{m}}(\gm)\, e^{i\gm\cdot\vec{r}}$ with $V_\mathrm{m}(-\mathbf{G}_\mathrm{m}) = V_\mathrm{m}(\mathbf{G}_\mathrm{m})^*$. The subleading corrections to $F_m$, quadratic in the moir\'e-scale momentum, give a momentum-dependent renormalization of $V_{\mathrm{m}}(\gm)$ that is small compared to the kinetic term and can be dropped.

\subsubsection{Full Hamiltonian}

Collecting the three terms, the full low-energy Hamiltonian is
\begin{equation}
\ham = \sum_{\vec{q}} \left\{ -\frac{q^2}{2m^*}\, c_{\vec{q}}^\dagger c_{\vec{q}}
+ \sum_{\gm} \left[ V_{\mathrm{m}}(\gm) + C_{ij}\, \epsilon_{ij}(\gm) + \left( \vec{q} + \frac{1}{2}\vec{G}_\mathrm{m} \right)_\a D_{\a ij}\, \epsilon_{ij}(\gm) \right] c_{\vec{q}+\gm}^\dagger c_{\vec{q}} \right\},
\label{eqn:hfull}
\end{equation}
where $q^2 = q_x^2 + q_y^2$, $V_{\mathrm{m}}(\gm)$ are the moir\'e potential Fourier coefficients, $C_{ij}$ and $D_{\a ij}$ are the scalar and pseudogauge strain couplings, and $\epsilon_{ij}(\gm)$ are the Fourier components of the strain tensor.

\subsection{From strain Hamiltonian to pseudomagnetic field}

We define the moir\'e-scale envelope function
\begin{equation}
\Psi(\vec{r}) = \frac{1}{\sqrt{|\Gamma_\mathrm{m}^*|}} \sum_{\vec{q}} e^{i\vec{q}\cdot\vec{r}}\, c_{\vec{q}}.
\end{equation}

To transform Eq.~\eqref{eqn:strain_q} into real space, we consider the $C_{ij}$ term and the momentum-dependent term, $D_{\a ij}$, separately. The $C_{ij}$ term gives
\begin{equation}
\sum_{\vec{q},\gm} C_{ij}\,\epsilon_{ij}(\gm)\, c_{\vec{q}+\gm}^\dagger c_{\vec{q}} = \int d^2\vec{r}\, \epsilon_{ij}(\vec{r})\, C_{ij}\, \Psi^\dagger(\vec{r})\, \Psi(\vec{r}).
\end{equation}
For the momentum-dependent term, the left-hand side carries $\left(\vec{q}+\frac{1}{2}\gm\right)_\a$. The $q_\a$ part becomes a spatial derivative through the integration by parts below, using $q_\a\, e^{i\vec{q}\cdot(\vec{r}'-\vec{r})} = -i\partial_{\vec{r}'}\, e^{i\vec{q}\cdot(\vec{r}'-\vec{r})}$ and $|\Gamma_\mathrm{m}^*|^{-1}\sum_{\vec{q}} (-i\partial_{\vec{r}'})\, e^{i\vec{q}\cdot(\vec{r}'-\vec{r})} = -i\partial_{\vec{r}'}\delta(\vec{r}'-\vec{r})$, while the $\frac{1}{2}\gm$ part transforms like the $C_{ij}$ term via $G_{\mathrm{m},\a}\,\epsilon_{ij}(\gm)\leftrightarrow -i\partial_\a\epsilon_{ij}(\vec{r})$, giving $-\frac{i}{2}\,\partial_\a\big(D_{\a ij}\epsilon_{ij}\big)\,\Psi^\dagger\Psi$. Together they give
\begin{equation}
\begin{split}
\sum_{\vec{q},\gm} \left(\vec{q}+\tfrac{1}{2}\gm\right)_\a D_{\a ij}\, \epsilon_{ij}(\gm)\, c_{\vec{q}+\gm}^\dagger c_{\vec{q}}
&= \int d^2\vec{r}\,\Psi^\dagger(\vec{r})\Big[-i\, D_{\a ij}\epsilon_{ij}(\vec{r})\,\partial_\a - \tfrac{i}{2}\,\partial_\a\big(D_{\a ij}\epsilon_{ij}(\vec{r})\big)\Big]\Psi(\vec{r}) \\
&= \int d^2\vec{r}\, \Psi^\dagger(\vec{r})\, \tfrac{1}{2}\left\{\hat k_\a,\, D_{\a ij}\epsilon_{ij}(\vec{r})\right\} \Psi(\vec{r}),
\end{split}
\end{equation}
where we eliminated the delta function derivative using integration by parts:
\begin{equation}
\Psi^\dagger(\vec{r}')\Psi(\vec{r})\,\partial_{\vec{r}'}\delta(\vec{r}'-\vec{r}) = -\partial_{\vec{r}'}\Psi^\dagger(\vec{r}')\,\Psi(\vec{r})\,\delta(\vec{r}'-\vec{r}).
\end{equation}
Define the scalar deformation potential $\Phi(\vec{r}) = C_{ij}\epsilon_{ij}(\vec{r})$ and the pseudogauge field
\begin{equation}
A_\a(\vec{r}) = \frac{1}{2}D_{\a ij}\,\epsilon_{ij}(\vec{r}),
\label{eqn:A_def}
\end{equation}
with units of energy and energy$\times$length, respectively, and $\hat k_\a \equiv -i\partial_\a$. The factor $\tfrac{1}{2}$ is included in the definition of $A_\a$ so that the pseudogauge Hamiltonian takes the standard anticommutator form, $\tfrac{1}{2}\left\{\hat k_\a, D_{\a ij}\epsilon_{ij}\right\} = \{\hat k_\a, A_\a\}$; note $D_{\a ij}\epsilon_{ij} = 2A_\a$. The strain Hamiltonian then takes the pseudogauge form
\begin{equation}
\ham_\epsilon = \int d^2\vec{r}\,\Psi^\dagger(\vec{r}) \left[\Phi(\vec{r}) + \{\hat k_\a, A_\a(\vec{r})\}\right]\Psi(\vec{r}).
\label{eqn:strain_gauge}
\end{equation}
The first term, $\Phi = C_{ij} \epsilon_{ij}$ is the scalar deformation potential that modifies the moir\'e potential, while the anticommutator represents the pseudomagnetic field contribution $\vec{B}(\vec{r}) = \nabla\times\vec{A}(\vec{r})$. The net pseudomagnetic flux through the moir\'e unit cell vanishes: by Stokes' theorem, $\Phi_B = \iint_S (\nabla\times\vec{A})\cdot d\vec{a} = \oint_{\partial S}\vec{A}\cdot d\vec{l}$, and contributions from opposite edges of the unit cell cancel pairwise since $\vec{A}(\vec{r}) = \vec{A}(\vec{r}+\vec{R})$ for any moir\'e lattice vector $\vec{R}$.

\section{Model parametrization}
\label{sec:param}

\subsection{Strain coupling coefficients}

For WSe$_2$/WS$_2$ heterostructures, the band structure near the valence band maximum is dominated by WSe$_2$ $d$ orbitals. It is therefore sufficient to use the tight-binding model for monolayer WSe$_2$~\cite{fang2018electronic}. The low-energy expansion of the unstrained and strain Hamiltonian in the two-band $(c, v)$ basis is
\begin{align}
\ham_\mathrm{unstrained,\,Fang} &= f_0 \hat{I} + \frac{f_1}{2}\hat{\sigma}_z + f_2 a (q_x \hat{\sigma}_x + q_y \hat{\sigma}_y), \label{eqn:hunstrained_fang} \\
\ham_{\epsilon,\,\mathrm{Fang}} &= f_3 \textstyle\sum_i \epsilon_{ii} + f_4 \textstyle\sum_i \epsilon_{ii} \hat{\sigma}_z + f_5[(\epsilon_{xx} - \epsilon_{yy})\hat{\sigma}_x - 2\epsilon_{xy}\hat{\sigma}_y], \label{eqn:strain_fang}
\end{align}
where $a$ is the monolayer lattice constant, $\hat{\sigma}_\a$ are Pauli matrices, and the parameters are $f_0 = -4.23$~eV, $f_1 = 1.65$~eV, $f_2 = 1.02$~eV, $f_3 = -5.26$~eV, $f_4 = -3.02$~eV, and $f_5 = 2.03$~eV.

To connect to Eq.~\eqref{eqn:strain_q}, we rewrite Eq.~\eqref{eqn:strain_fang} in terms of $\epsilon_0 = \epsilon_{xx} + \epsilon_{yy}$, $\epsilon_1 = \epsilon_{xx} - \epsilon_{yy}$, $\epsilon_2 = -2\epsilon_{xy}$, and Fourier-transform:
\begin{equation}
\ham_{\epsilon,\,\mathrm{Fang}} = \sum_{\vec{q},\gm} \left\{ (f_3 \hat{I} + f_4 \hat{\sigma}_z)\, \epsilon_0(\gm) + f_5 \left[ \epsilon_1(\gm)\, \hat{\sigma}_x + \epsilon_2(\gm)\, \hat{\sigma}_y \right] \right\} c_{\vec{q}+\gm}^\dagger c_{\vec{q}}.
\label{eqn:strain_fang_ft}
\end{equation}
Promoting $C_{ij}$ and $D_{\a ij}$ to $2\times 2$ matrices $\hat{C}_{ij}$ and $\hat{D}_{\a ij}$ and comparing with Eq.~\eqref{eqn:strain_fang_ft}, we find
\begin{align}
\hat{C}_{xx} &= f_3 \hat{I} + f_4 \hat{\sigma}_z + f_5 \hat{\sigma}_x, \nonumber \\
\hat{C}_{yy} &= f_3 \hat{I} + f_4 \hat{\sigma}_z - f_5 \hat{\sigma}_x, \nonumber \\
\hat{C}_{xy} &= -f_5 \hat{\sigma}_y. \label{eqn:c_coeff}
\end{align}
At the two-band level, Fang's $\ham_{\epsilon,\,\mathrm{Fang}}$ contains no $q_\a \epsilon_{ij}$ cross term, so the bare $\hat{D}_{\a ij}$ vanish. A nonzero $D_{\a ij}$ in the valence-band sector is generated by the Schrieffer-Wolff projection.

Since the band gap $f_1 = 1.65$~eV is much larger than the moir\'e bandwidth, we project onto the valence band:
\begin{equation}
\ham_\mathrm{eff} = H_{vv} - \frac{H_{vc}\, H_{cv}}{E_c - E_v},
\end{equation}
with $H_{vv} = (f_3 - f_4)\,\epsilon_0$ and
\begin{equation}
H_{vc} = f_2 a (q_x - iq_y) + f_5 (\epsilon_1 - i\epsilon_2), \quad H_{cv} = H_{vc}^*.
\end{equation}
The effective Hamiltonian is
\begin{align}
\ham_\mathrm{eff} &= (f_3 - f_4)\,\epsilon_0 - \frac{1}{f_1} \left[ f_2^2 a^2 q^2 + 2 f_2 f_5 a (\epsilon_1 q_x + \epsilon_2 q_y) + f_5^2 (\epsilon_1^2 + \epsilon_2^2) \right] \nonumber \\
&= (f_3 - f_4)\,\epsilon_0 - \frac{|f_2 a \vec{q} + f_5 \vec{\epsilon}|^2}{f_1},
\label{eqn:heff}
\end{align}
where $\vec{q} = (q_x, q_y)$ and $\vec{\epsilon} = (\epsilon_1, \epsilon_2)$. The parabolic term gives the effective mass $m^* = \hbar^2 f_1 / (2 f_2^2 a^2) = 0.55\, m_0$, using $a = 3.316$~\AA\ from Ref.~\cite{fang2018electronic}. From the cross term, the projected strain coefficients are
\begin{align}
C_{xx} = C_{yy} &= f_3 - f_4 = -2.25~\text{eV}, \nonumber \\
D_{xxx} = -\mathcal{D}, \quad D_{xyy} &= \mathcal{D}, \quad D_{yxy} = D_{yyx} = \mathcal{D},
\end{align}
where $\mathcal{D} \equiv 2 f_2 f_5 a / f_1 = 8.32$~eV$\cdot$\AA. The projected single-band strain Hamiltonian is
\begin{equation}
\ham_\epsilon(\vec{q}) = \sum_{\gm} \left[ (f_3 - f_4)\, \epsilon_0(\gm) - \frac{2 f_2 f_5 a}{f_1} \left( \epsilon_1(\gm)\, q_x + \epsilon_2(\gm)\, q_y \right) \right] c_{\vec{q}+\gm}^\dagger c_{\vec{q}}.
\label{eqn:strain_final}
\end{equation}
Here the $\vec{q}$-linear term is written with the classical symbol $\vec{q}$; Hermitian symmetrization replaces $\vec{q} \to \vec{q} + \gm/2$, reproducing the midpoint momentum of Eq.~\eqref{eqn:strain_q} and, in real space, the anticommutator form of Eq.~\eqref{eqn:strain_gauge}. We neglect the term $-f_5^2(\epsilon_1^2+\epsilon_2^2)/f_1$ in Eq.~\eqref{eqn:heff}, quadratic in strain, which reaches only ${\sim}0.5$~meV at the domain walls and acts as a small additional scalar potential.
Combining with Eq.~\eqref{eqn:A_def}, the pseudogauge field for the specific WSe$_2$/WS$_2$ system is
\begin{equation}
A_\a(\vec{r}) = \frac{1}{2}D_{\a ij}\,\epsilon_{ij}(\vec{r}) = -\frac{1}{2}\mathcal{D}\, \epsilon_\a(\vec{r}) = -\frac{f_2 f_5 a}{f_1}\,\epsilon_\a(\vec{r}), \qquad \frac{f_2 f_5 a}{f_1} = 4.16~\text{eV}\cdot\text{\AA},
\label{eqn:A_explicit}
\end{equation}
matching the definition used in the main text [Eq.~(5)]. Note that even though the moir\'e potential and relaxation pattern have $C_3$ symmetry, the individual strain Fourier components are not necessarily related by $C_3$, so all coefficients must be retained.

\subsection{Moir\'e potential}
\label{sec:parametrization}

\subsubsection{Stacking-dependent potential and connection to DFT}

In the local configuration approximation~\cite{carr2018relaxation,cazeaux2020energy,relaxfosdick22}, the moir\'e potential at position $\vec{r}$ is determined by the local stacking vector
\begin{equation}
\vec{b}(\vec{r}) = (M - \mathbb{I})\,\vec{r}, \qquad M = A_1 A_2^{-1} = R(\theta/2)\, S_1\, R(-\theta/2)\, S_2^{-1},
\end{equation}
where $R(\theta)$ is a rotation matrix and $S_{1,2}$ encode each layer's lattice constant. We take layer~1 = WS$_2$ (unstretched, $S_1 = \mathbb{I}$) and layer~2 = WSe$_2$ (stretched, $S_2 = (1+\delta)\mathbb{I}$). Since $|M - \mathbb{I}| \ll 1$, the stacking $\vec{b}(\vec{r})$ varies slowly on the monolayer scale. In this limit, the moir\'e potential is $V_{\mathrm{m}}(\vec{r}) = \mathcal{V}(\vec{b}(\vec{r}))$, where $\mathcal{V}(\vec{b})$ is the stacking-dependent valence band edge computed by DFT on rigid commensurate bilayers at each shift $\vec{b}$.

The function $\mathcal{V}(\vec{b})$ is periodic with the monolayer unit cell, so we expand in monolayer reciprocal lattice vectors $\vec{G}$:
\begin{equation}
\mathcal{V}(\vec{b}) = \sum_{\vec{G}} \mathcal{V}(\vec{G})\, e^{i\vec{G}\cdot\vec{b}}, \qquad
\mathcal{V}(\vec{G}) = \frac{1}{|\Omega_\mathrm{uc}|} \int_{\Omega_\mathrm{uc}} d^2\vec{b}\, \mathcal{V}(\vec{b})\, e^{-i\vec{G}\cdot\vec{b}}.
\end{equation}
Substituting $\vec{b}(\vec{r}) = (M - \mathbb{I})\vec{r}$ gives
\begin{equation}
V_{\mathrm{m}}(\vec{r}) = \sum_{\vec{G}} \mathcal{V}(\vec{G})\, e^{i\vec{G}\cdot(M-\mathbb{I})\vec{r}} = \sum_{\gm} V_{\mathrm{m}}(\gm)\, e^{i\gm\cdot\vec{r}},
\end{equation}
with the identification
\begin{equation}
\gm = (M - \mathbb{I})^\top\, \vec{G}, \qquad V_{\mathrm{m}}(\gm) = \mathcal{V}(\vec{G}).
\label{eqn:vgm_dft}
\end{equation}
Thus, the moir\'e reciprocal lattice is the image of the monolayer reciprocal lattice under the small linear map $(M - \mathbb{I})^\top$, and the Fourier coefficients $V_{\mathrm{m}}(\gm)$ are the monolayer Fourier components of the DFT-computed stacking energy. Because $\mathcal{V}(\vec{b})$ is smooth, in the unrelaxed limit we truncate $\vec{G}$ to the first few shells~\cite{wu2018hubbard}.

\subsubsection{DFT calculations and rigid moir\'e potential}

\begin{figure}[h]
    \centering
    \includegraphics[width=0.6\linewidth]{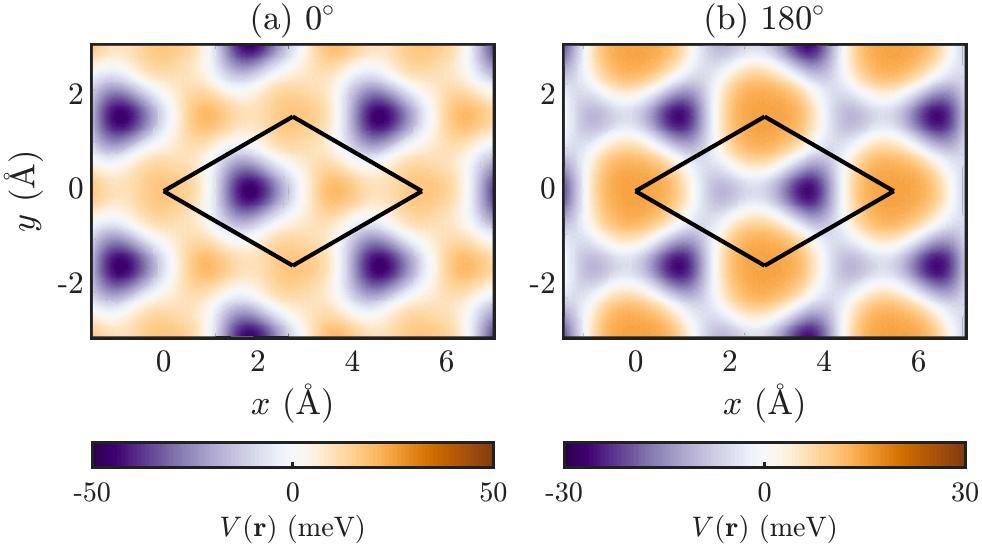}
    \caption{Fitted rigid moir\'e potential for (a) 0$^\circ$ and (b) 180$^\circ$ WSe$_2$/WS$_2$.}
    \label{fig:rigid_potential}
\end{figure}

To obtain $\mathcal{V}(\vec{b})$, we compute the valence band edge $E_v(\vec{b})$ at each stacking vector $\vec{b}$ and define $V_\mathrm{m}(\vec{b})$ as the deviation from the mean: $V_\mathrm{m}(\vec{b}) = E_v(\vec{b}) - \sum_{\vec{b}} E_v(\vec{b})/N$, where $N$ is the number of discrete shifts. The band edge is calculated with respect to the reference vacuum energy.
We sample a $9\times9$ grid in stacking space and fit to the Fourier form
\begin{equation}
V_\mathrm{m}(\vec{b}) = \sum_s \sum_j 2V_s \cos(\vec{G}_j^s \cdot \vec{b} + \phi_s),
\label{eq:app_V_rigid}
\end{equation}
where $s$ is the reciprocal-lattice shell index and $j$ runs over the $C_3$-symmetric vectors in each shell. This functional form ensures that $\int d^2\vec{b}\, V_\mathrm{m}(\vec{b}) = 0$. The fitted coefficients are given in Table~\ref{tab:coeff}, and the resulting real-space potential is shown in Fig.~\ref{fig:rigid_potential}.

\begin{table}[h]
\centering
\begin{tabular}{|c| c| c| c| c|}
  \hline
 & \multicolumn{2}{c|}{SCAN-rVV10} & \multicolumn{2}{c|}{DFT-D3} \\
\hline
Shell $s$ & $V_s$ (meV) &  $\phi_s$ ($^\circ$) & $V_s$ (meV) & $\phi_s$ ($^\circ$) \\
\hline
1 & 3.659   & 50.89 & 4.902 & 53.97 \\
2 & $-$0.279 & 10.60 & $-$0.303 & $-$0.155 \\
3 & 0.603  & $-$110.2 & 0.716 & $-$98.43 \\
\hline
\end{tabular}
\caption{Fitted moir\'e potential Fourier coefficients by reciprocal-lattice shell for two van der Waals functionals.}
\label{tab:coeff}
\end{table}

To generate the reciprocal lattice vectors for a given shell, we define the reciprocal basis vectors $\vec{G}_1$ and $\vec{G}_2$ from the real-space lattice matrix $\mathbf{A}$ via $\mathbf{G} = 2\pi(\mathbf{A}^{-1})^T$. For a hexagonal lattice, vectors are grouped into shells based on their squared magnitude, which is proportional to the metric $Q(m, n) = m^2 + n^2 - mn$ for a vector $\vec{G} = m\vec{G}_1 + n\vec{G}_2$. We identify shell radii from the distinct sorted values of $Q$, solve the Diophantine equation $m^2 + n^2 - mn = Q_s$ for each shell $s$ to find all integer pairs $(m, n)$, and construct the physical vectors by linear combination.

\subsubsection{Relaxed moir\'e potential}

\begin{figure*}
    \centering
    \includegraphics[width=\linewidth]{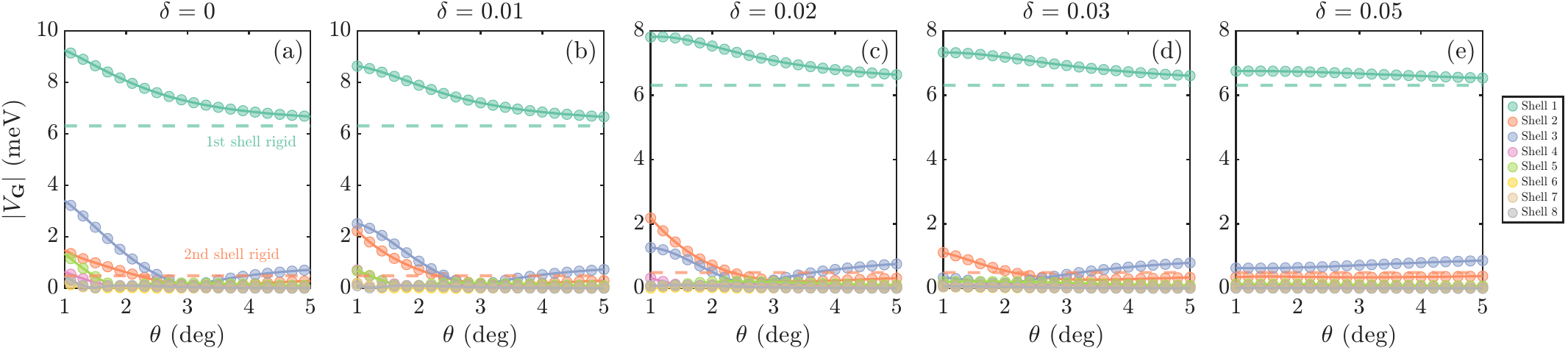}
    \caption{Fourier shell decomposition of the relaxed moir\'e potential as a function of twist angle $\theta$ for several values of lattice mismatch: (a) $\delta = 0$, (b) $\delta = 0.01$, (c) $\delta = 0.02$, (d) $\delta = 0.03$, (e) $\delta = 0.05$. Each panel shows the magnitude of the Fourier coefficient $|V_j^{(s)}|$ for shells $s = 1$ through 8 (out of 15 used in the calculation). Dashed horizontal lines indicate the first- and second-shell coefficients of the rigid potential. At large $\theta$, the relaxed coefficients approach the rigid values. At small $\theta$, higher shells acquire significant weight due to the sharpening of domain walls.}
    \label{fig:V_vs_theta}
\end{figure*}

With relaxation, the interlayer bond connecting atoms in layers 1 and 2 is shifted by the displacement difference $\Delta\vec{u}(\vec{r}) = \vec{u}_1(\vec{r}) - \vec{u}_2(\vec{r})$ at zeroth order in the bond length (unlike the intralayer case, where both atoms share the same displacement field and the leading correction involves the gradient $\nabla\vec{u}$). The local stacking is therefore shifted to $\vec{b}(\vec{r}) + \Delta\vec{u}(\vec{r})$, giving
\begin{equation}
V_{\mathrm{m,relax}}(\vec{r}) = \sum_{\vec{G}} \mathcal{V}(\vec{G})\, e^{i\vec{G}\cdot\vec{b}(\vec{r})}\, e^{i\vec{G}\cdot\Delta\vec{u}(\vec{r})}
= \sum_s \sum_j 2V_s \cos\!\left[\vec{G}_j^s \cdot \left(\vec{b}(\vec{r}) + \Delta\vec{u}(\vec{r})\right) + \phi_s\right].
\label{eqn:V_relax}
\end{equation}
To parametrize the relaxed potential, we adopt a general Fourier decomposition:
\begin{equation}
V_\mathrm{m,\,relax}(\vec{r}) = \sum_{\gm} C_{\gm}\, e^{i\gm \cdot \vec{r}},
\end{equation}
where the reality of the potential imposes $C_{\gm}^* = C_{-\gm}$. We use this representation rather than re-fitting to the shell form for two reasons. First, the sharpening of domain walls by atomic reconstruction generates significant weight in higher Fourier shells: as shown in Fig.~\ref{fig:V_vs_theta}, shells beyond the leading two acquire amplitudes comparable to the first shell at small twist angles, and the coefficients do not decay rapidly with shell index. Second, the shell-based fitting ansatz assumes that all coefficients within a given shell are identical, which is not guaranteed for the relaxed potential. Fitting to the shell form can therefore produce Gibbs-like oscillations within the domains, whereas the direct Fourier decomposition is exact and requires no fitting.

Figure~\ref{fig:V_vs_theta} shows the magnitude of the Fourier coefficients for shells 1 through 8 as a function of twist angle $\theta$ at several values of lattice mismatch $\delta$. At large $\theta$ (weak relaxation), the coefficients approach the rigid-model values (dashed lines), with the first shell dominant and higher shells negligible. As $\theta$ decreases, relaxation sharpens the domain walls and the higher-shell coefficients grow, with shells 3--8 reaching amplitudes on the order of 1~meV. In the calculation, we retain up to 15 shells to ensure convergence; the first 8 are shown in the figure.

\subsection{Unit conversion for pseudomagnetic field}

The vector potential in the code has units of $[\vec{A}_\mathrm{code}] = \mathrm{eV\cdot\text\AA}$. Since the Hamiltonian with a physical vector potential takes the form
\begin{equation}
\ham = \frac{(\hbar\vec{q} + e\vec{A}_\mathrm{phys})^2}{2m^*} \approx \frac{\hbar^2 q^2}{2m^*} + \frac{e\hbar}{m^*}\vec{A}_\mathrm{phys}\cdot\vec{q},
\label{eq:app_A_relax}
\end{equation}
we identify $\vec{A}_\mathrm{phys} = (2m^*/e\hbar)\,\vec{A}_\mathrm{code}$. (The hole band carries an overall minus sign on the minimal-coupling form, which flips the sign but not the magnitude of $\vec{A}_\mathrm{phys}$ and $B_z$.) Writing $m^* = M m_e$ with $M = 0.55$ and using $m_e/\hbar = 8.634\times10^3$~s/m$^2$, the pseudomagnetic field in physical units is obtained by taking the curl and converting from \AA$^{-1}$ to m$^{-1}$.

\subsection{Computational details}
\label{sec:computational}

The Hamiltonian is represented in a plane-wave basis of moir\'e reciprocal lattice vectors $\vec{q} + \gm$, where $\gm = m\vec{g}_1 + n\vec{g}_2$ ranges over all moir\'e reciprocal lattice vectors within a circular cutoff. We perform a finite truncation including 163 basis elements. Band structures are computed along high-symmetry paths $\bar{\Gamma}$--$\bar{M}$--$\bar{K}$--$\bar{\Gamma}$ with 80 $k$-points per segment. Chern numbers are evaluated on a $31\times31$ $k$-grid, the Berry curvature and quantum metric on a finer $61\times61$ grid, and the trace-condition diagnostics $T/2\pi$ and $F$ (Fig.~4 of the main text) on a $41\times41$ grid; the corresponding quantum-geometric definitions are given in Sec.~\ref{sec:quantum-geometry}. The phase diagrams are computed on an $81 \times 21$ grid in $(\theta, \delta)$ space, with $\theta \in [0^\circ, 2^\circ]$ in steps of $0.025^\circ$ and $\delta \in [0, 0.05]$ in steps of $0.0025$. The moir\'e potential and strain coefficients at each $(\theta, \delta)$ point are obtained by bilinear interpolation from the precomputed relaxation grid when the target point does not coincide with a grid node.

\subsection{Quantum geometry and superfluid stiffness}
\label{sec:quantum-geometry}

Quantum-geometric quantities are extracted from the cell-periodic Bloch states $\ket{u_n(\vec{k})}$ of the projected valence-band Hamiltonian on a uniform $N\times N$ moir\'e Brillouin-zone mesh, $\vec{k}=k_1\vec{b}_1+k_2\vec{b}_2$ with $k_{1,2}\in\{0,1/N,\dots,(N-1)/N\}$. All quantities are built from the gauge-invariant overlaps between neighboring states,
\begin{equation}
P_\mu(\vec{k})=\braket{u_n(\vec{k})}{u_n(\vec{k}+\Delta\vec{k}_\mu)},
\end{equation}
where $\Delta\vec{k}_1,\Delta\vec{k}_2$ are the two primitive mesh steps and $\Delta\vec{k}_{12}=\Delta\vec{k}_1+\Delta\vec{k}_2$ is the plaquette diagonal.

The Berry curvature is obtained from the discrete link variables $U_\mu(\vec{k})=P_\mu(\vec{k})/|P_\mu(\vec{k})|$ following Ref.~\cite{fukui2005chern}. The lattice field strength on the mesh plaquette anchored at $\vec{k}$ is
\begin{equation}
\tilde{F}_{12}(\vec{k})=\arg\!\left[U_1(\vec{k})\,U_2(\vec{k}+\Delta\vec{k}_1)\,U_1^{-1}(\vec{k}+\Delta\vec{k}_2)\,U_2^{-1}(\vec{k})\right],
\end{equation}
which gives the Chern number $C=(2\pi)^{-1}\sum_{\vec{k}}\tilde{F}_{12}(\vec{k})$ and the local Berry curvature $\Omega(\vec{k})=\tilde{F}_{12}(\vec{k})/A_{\mathrm{p}}$, where $A_{\mathrm{p}}=|\Gamma^*_\mathrm{m}|/N^2$ is the area of one mesh plaquette. The quantum (Fubini--Study) metric follows from the quantum distance between neighboring states, $1-|P_\mu|^2=g_{\mu\nu}\,\Delta k^\mu\Delta k^\nu+\mathcal{O}(\Delta k^3)$, evaluated along the two primitive directions and along the diagonal:
\begin{equation}
\begin{aligned}
g_{11}&=N^2\!\left(1-|P_1|^2\right),\qquad g_{22}=N^2\!\left(1-|P_2|^2\right),\\
g_{12}&=\tfrac{1}{2}N^2\!\left[\left(1-|P_{12}|^2\right)-\left(1-|P_1|^2\right)-\left(1-|P_2|^2\right)\right].
\end{aligned}
\end{equation}
These dimensionless lattice-basis components are rotated into the Cartesian frame through $g^{\mathrm{cart}}=W^{\!\top}g\,W$, with $W=[\vec{b}_1\ \vec{b}_2]^{-1}$ the inverse of the matrix of moir\'e reciprocal vectors.

The proximity of a band to ideal (K\"ahler) quantum geometry is quantified by the trace-condition violation $T(\vec{k})=\tr g(\vec{k})-|\Omega(\vec{k})|\geq0$, with integrated value $T/2\pi=(2\pi)^{-1}\!\int_{\mathrm{mBZ}}\!\left(\tr g-|\Omega|\right)d^2k$ (vanishing in the ideal limit), and by the Berry-curvature fluctuation $F=\sqrt{\langle(\tilde\Omega-C)^2\rangle_{\mathrm{mBZ}}}$, where $\tilde\Omega(\vec{k})=\Omega(\vec{k})\,|\Gamma^*_\mathrm{m}|/(2\pi)$ is the dimensionless per-cell Berry curvature normalized so that $\langle\tilde\Omega\rangle_{\mathrm{mBZ}}=C$, and $F$ remains well defined for trivial bands ($C=0$).

The superfluid stiffness tensor is the Cartesian quantum metric integrated over the moir\'e Brillouin zone, $D_{s,\a\b}=|\Gamma^*_\mathrm{m}|^{-1}\!\int_{\mathrm{mBZ}}g^{\mathrm{cart}}_{\a\b}(\vec{k})\,d^2k$, and $\bar{D}_s=\tfrac{1}{2}(\lambda_++\lambda_-)$ is the mean of its two eigenvalues $\lambda_\pm$. Berry curvature, quantum metric, and trace diagnostics are evaluated on the $k$-meshes quoted above.


\section{Continuum lattice relaxation model}
\label{sec:relaxation}

\subsection{Lattice relaxation}

The equilibrium displacement field is obtained by minimizing the total energy within a continuum configuration-space framework~\cite{cazeaux2020energy, carr2018relaxation}. The total energy per moir\'e unit cell consists of two contributions:
\begin{equation}
E[\vec{u}_1, \vec{u}_2] = E_\mathrm{el}[\vec{u}_1, \vec{u}_2] + E_\mathrm{misfit}[\vec{u}_1, \vec{u}_2],
\end{equation}
where $\vec{u}_\ell(\vec{r})$ is the in-plane displacement field of layer $\ell = 1, 2$ (layer~1 = WS$_2$, layer~2 = WSe$_2$, following the convention of Sec.~\ref{sec:derivation}).

The elastic energy is
\begin{equation}
E_\mathrm{el} = \frac{1}{2} \sum_{\ell=1}^{2} \int_\Omega C_{ijkl}^{(\ell)}\, \epsilon_{ij}^{(\ell)}(\vec{r})\, \epsilon_{kl}^{(\ell)}(\vec{r})\, d^2\vec{r},
\end{equation}
where $\epsilon_{ij}^{(\ell)} = \tfrac{1}{2}(\partial_i u_j^{(\ell)} + \partial_j u_i^{(\ell)})$ is the linearized strain tensor of layer $\ell$, and $C_{ijkl}^{(\ell)}$ is the elastic tensor parameterized by the bulk modulus $K_\ell$ and shear modulus $G_\ell$. For WS$_2$ (layer~1) we use $K_1 = 53.37$~eV/\AA$^2$, $G_1 = 33.23$~eV/\AA$^2$; for WSe$_2$ (layer~2), $K_2 = 46.87$~eV/\AA$^2$, $G_2 = 31.21$~eV/\AA$^2$, obtained from stretching and shearing the monolayers.

The misfit energy penalizes unfavorable local stacking and is given by the GSFE evaluated at the local stacking vector:
\begin{equation}
E_\mathrm{misfit} = \int_\Omega V_\mathrm{GSFE}\!\left(\vec{b}(\vec{r}) + \Delta\vec{u}(\vec{r})\right) d^2\vec{r},
\end{equation}
where $\Delta\vec{u}(\vec{r}) = \vec{u}_1(\vec{r}) - \vec{u}_2(\vec{r})$ is the relative displacement and $\vec{b}(\vec{r})$ is the rigid stacking vector. Following the convention of Sec.~\ref{sec:derivation}, layer~1 (WS$_2$) is unstretched ($S_1 = \mathbb{I}$) and layer~2 (WSe$_2$) carries the lattice mismatch ($S_2 = (1+\delta)\mathbb{I}$). The local stacking vector is
\begin{equation}
\vec{b}(\vec{r}) = (M - \mathbb{I})\,\vec{r}, \qquad M = R(\theta/2)\, S_1\, R(-\theta/2)\, S_2^{-1},
\end{equation}
which varies linearly across space. The moir\'e superlattice vectors are $\mathbf{A}_\mathrm{sc} = (M - \mathbb{I})^{-1}\, \mathbf{A}$. As $\vec{r}$ traverses one moir\'e unit cell, $\vec{b}(\vec{r})$ sweeps through the full monolayer unit cell exactly once, establishing a one-to-one map between real space and configuration (stacking) space parameterized by the fractional coordinates $(s, t) = \mathbf{A}^{-1}\vec{b}$.

\subsection{Generalized stacking fault energy}

\begin{figure}[h]
    \centering
    \includegraphics[width=0.55\linewidth]{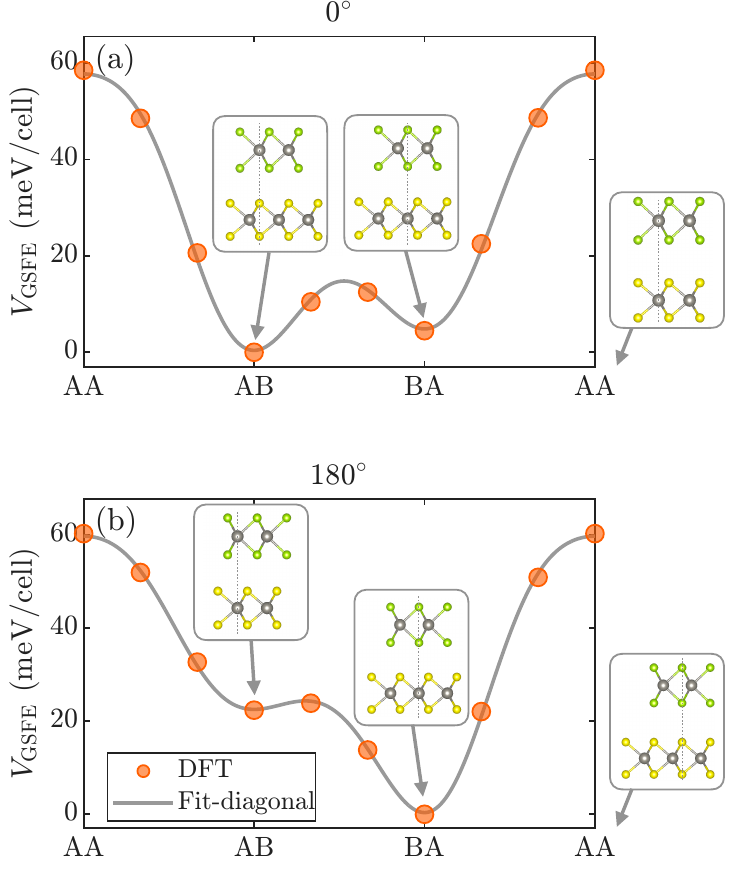}
    \caption{GSFE for (a) 0$^\circ$ and (b) 180$^\circ$ WSe$_2$/WS$_2$ obtained using the SCAN-rVV10 functional. Points: DFT; lines: fitted model. High-symmetry stackings are shown in insets.}
    \label{fig:gsfe}
\end{figure}

The GSFE is calculated for aligned WSe$_2$/WS$_2$ bilayers at a series of in-plane shifts using both SCAN-rVV10 and DFT-D3 functionals. We use a $\Gamma$-centered $k$-mesh of $12\times12\times1$ and a plane-wave cutoff of 500~eV. The interlayer distance is relaxed at each stacking while keeping atomic positions within each layer fixed. Figure~\ref{fig:gsfe} shows the GSFE for both 0$^\circ$ and 180$^\circ$ stacking configurations. The SCAN-rVV10 functional, being a fully nonlocal correlation functional, is expected to be more accurate than the semi-empirical DFT-D3 dispersion scheme. The two functionals produce qualitatively similar GSFE landscapes. 

The GSFE is parameterized by a Fourier expansion in the fractional stacking coordinates $(s, t) = \mathbf{A}^{-1}\vec{b}$, where $\mathbf{A}$ is the monolayer lattice matrix and $\vec{b}$ is the local stacking vector:
\begin{equation}
\begin{aligned}
V_\mathrm{GSFE}(s, t) &= c_1\left(\cos s + \cos t + \cos(s+t)\right) \\
&+ c_2\left(\cos(s+2t) + \cos(s-t) + \cos(2s+t)\right) \\
&+ c_3\left(\cos 2s + \cos 2t + \cos 2(s+t)\right) \\
&+ c_4\left(\sin s + \sin t - \sin(s+t)\right) \\
&+ c_5\left(\sin(2s+2t) - \sin 2s - \sin 2t\right).
\end{aligned}
\end{equation}
The symmetric terms ($c_1$--$c_3$) respect the full $C_6$ symmetry of the hexagonal stacking space, while the antisymmetric terms ($c_4$, $c_5$) break the equivalence between AB and BA stackings. The coefficients are fitted to the DFT data in Fig.~\ref{fig:gsfe}. For the 3R (0$^\circ$) configuration, the GSFE has two inequivalent minima at the AB and BA stackings, with AB being the global minimum; the antisymmetric coefficients ($c_4$, $c_5$) break the degeneracy between the two. For 2H (180$^\circ$), the global minimum is at the AA stacking, and the asymmetry between AB and BA is much weaker ($c_4 = +0.70$~meV vs.\ $c_4 = -3.68$~meV for 3R).

The energy is discretized on a periodic $N \times N$ grid ($N = 36$) in the hull (configuration space), with the elastic term evaluated in Fourier space via FFT. The total energy is minimized using the L-BFGS algorithm to obtain the equilibrium displacements $\vec{u}_1$, $\vec{u}_2$. The strain tensor $\epsilon_{ij}(\vec{r})$ used in the continuum electronic model is computed from the relative displacement $\Delta\vec{u} = \vec{u}_1 - \vec{u}_2$ via spectral differentiation.

\section{Computational details of neural network calculations}
\label{sec:nnvmc}

{\em Neural Network Wavefunction and Variational Monte Carlo.---}
To solve the continuum Schr\"odinger equation, we minimize the energy directly over a neural-network parameterization of the many-electron wavefunction within variational Monte Carlo.
We write the wavefunction as a sum of $K$ determinants built from generalized orbitals,
\begin{equation}
\Psi(\bm x_1, \ldots, \bm x_N) = \sum_{k = 1}^K \det_{ij} \phi_{j}^k(\bm x_i, {\bm x_{/i}}),
\label{eq:1}
\end{equation}
where $\bm x_i = (\bm r_i,\sigma_i)$ collects the position $\bm r_i$ and spin $\sigma_i$ of electron $i$ \cite{Pfau2020Sep,vonGlehn2022Nov,Geier2025Attn}.
Many-body correlations enter explicitly through the generalized orbitals $\phi_j^k(\bm x_i;{\bm x_{/i}})$, whose dependence on the configuration of all remaining electrons makes the orbital of electron $i$ conditioned on its environment.
Crucially, this construction endows Eq.~\eqref{eq:1} with an amplitude and nodal structure flexible enough to \emph{universally} represent two-dimensional fermionic wavefunctions with only a handful of determinants \cite{fu2025fermions,fu2026fermisets}. Because electron coordinate and spin are embedded jointly and on equal footing, the generalized orbitals retain their general spinful form \cite{avdoshkin2025spin}.
We realize the orbitals in Eq.~\eqref{eq:1} with a transformer network, a universal approximator of permutation-equivariant sequence-to-sequence maps ($\{ \bm x_i\} \rightarrow \{\phi_j \}$) \cite{Yun2020TransformerUniversalApprox}.
Refs.~\cite{vonGlehn2022Nov,Geier2025Attn} give the full construction of the neural-network wavefunction.
Energy expectation values and observables are evaluated efficiently by Markov-chain Monte-Carlo sampling \cite{Foulkes2001Jan}.
For the parameter updates we adopt a Kronecker-factored curvature approximation \cite{Martens2015Jun}, which yields faster and more stable convergence \cite{Pfau2020Sep,ChenJ1J22024,Goldshlager2024}.
The variational Monte Carlo method is described in detail in Refs.~\cite{Foulkes2001Jan,Pfau2020Sep,Geier2025Attn}, while aspects specific to spinful systems are described in Refs.~\cite{Schmidt1999,Ambrosetti2012,Melton2016QMCspin,Melton2016spinorbit,adams2021variational,gerard2024solids,luo2025solving,avdoshkin2025spin}.

The hyperparameters that were used in training the neural network wavefunction with VMC are contained in Tab.~\ref{tab:hyperparameters}.
The neural network VMC implementation used in this work is based on Refs.~\cite{Pfau2020Sep,vonGlehn2022Nov,Geier2025Attn} and our code is publicly available in our codebase {\tt PeriodicWave} \cite{periodicwave_github} with implementation details in Ref.~\cite{Geier2025Attn}.

The Hamiltonian solved in the neural network calculation directly uses the rigid [Eq.~\eqref{eq:app_V_rigid}] or relaxed moire potential [Eqs.~\eqref{eqn:V_relax} and \eqref{eq:app_A_relax}] together with Coulomb interaction energy $\frac{e^2}{4 \pi \epsilon \epsilon_0 |\bm r_i - \bm r_j|}$ between electrons $i,\ j$ with relative dielectric constant $\epsilon = 5$.
The rigid and relaxed calculations use identical wavefunction architecture, optimization protocol, and interaction, and differ only in the single-particle input: the rigid calculation uses the rigid moir\'e potential, while the relaxed calculation uses the relaxed moir\'e potential together with the strain-induced pseudoelectric and pseudogauge terms. 
The twist angle is swept from $2^\circ$ to $2.8^\circ$ at fixed lattice mismatch $\delta = 0.02$. Error bars denote the Monte Carlo standard errors of the three total energies propagated to $E_c$.

{\em Charge gap.---}
The many-body charge gap shown in Fig.~1(e)-(f) of the main text is defined as
\begin{equation}
E_c = E(N+1) + E(N-1) - 2E(N),
\end{equation}
where $E(N)$ is the NN-VMC ground-state energy of the $N$-particle system. This quantity is a measure of the band gap that is invariant under gauge transformations introducing a constant energy shift $\hat H \to \hat H + \sum_i^N \mu \hat c^\dagger_i \hat c_0$. Thereby, the subtraction of the homogeneous charge background that must to be performed in the calculation of Coulomb energy \cite{Geier2025Attn} cancels exactly. The bare energies $E(N-1),\, E(N),\, E(N+1)$ from which the charge gap is computed are reported in Table~\ref{tab:nnvmc_energies}.

{\em Finite size effects.---}
We use $N = 18$ electrons in 9 moir\'e unit cells in a triangular $3 \times 3$ periodic supercell, corresponding to filling factor $\nu = 2$. The $3 \times 3$ supercell includes the $K$ point but neglects the $M$ point.
Because the single-particle band structure becomes metallic with the $K$ point of the topmost band reaching below the $M$ point of the maximum of the second most band at twist $\theta \geq 2.8^\circ$ for the rigid and $\theta \geq 3.3^\circ$ for the relaxed model, we restrict the many-body calculations to $\theta \leq 2.8^\circ$. 
We expect that for $\theta > 2.8^\circ$, neglecting the $M$ point in the finite size of the $3 \times 3$ geometry would neglect relevant effects that may yield the numerical estimate unreliable. 

{\em Ground state phase.---} 
In the range $\theta = 2^\circ$ to $2.8^\circ$, the spin-density extracted from the obtained ground state wavefunction indicates that the system is in a paramagnetic phase. 
Below $\theta < 2^\circ$, the spin density calculations indicate the appearance of an antiferromagnetic phase for both the relaxed and rigid model. 
Since the magnetic phase introduces novel physical effects that affect the interpretation of the charge gap, we restrict the discussion to $\theta \geq 2^\circ$.
Magnetism in moir\'e materials obtained from NNVMC calculations has been discussed recently in Ref.~\cite{geier2026magnetism}.

\begin{table}
    \centering
    \renewcommand{\arraystretch}{1.2}
    \begin{tabular}{lll}
        \hline\hline
        \textbf{Category} & \textbf{Parameter} & \textbf{Value} \\
        \hline
        \multirow{8}{*}{Architecture} & Network layers & $L = 2$ \\
        & Attention heads per layer & $N_{\rm heads} = 4$ \\
        & Attention dimension & $d_{\rm k} = 32$ \\
        & Value dimension & $d_{\rm v} = 32$ \\
        & Perceptron dimension & $d_{\rm model} = 128$ \\
        & $\#$ perceptrons per layer & $2$ \\
        & Determinants & $K = 4$ \\
        & Layer norm & True \\
        \hline
        \multirow{5}{*}{Training} & Learning rate schedule & $\eta = \eta_0 \frac{2 t}{t_d} \frac{1}{1 + (t/t_d)^2}$ \\
        & Local energy clipping & $\rho = 5.0$ \\
        & Training iterations & $300{,}000$ \\
        & Learning rate & $\eta_0 = 0.03$ \\
        & Learning rate delay & $t_0 = 3000$ \\
        \hline
        \multirow{2}{*}{MCMC} & Batch size & $4096$ \\
        & Initial move width & $0.1$ \\
        \hline
        \multirow{2}{*}{KFAC} & Norm constraint & $1 \times 10^{-3}$ \\
        & Damping & $1 \times 10^{-4}$ \\
        \hline
         \multirow{2}{*}{Fine tuning} & iterations & $27,000$ \\
        & Batch size & $16,384$ \\
        \hline\hline
    \end{tabular}
    \caption{Default hyperparameters used in the NNVMC calculations based on the self-attention neural network. The training was performed in two stages: 300,000 training steps at batch size $4096$ were followed by 27,000 fine-tuning steps at batch size $16384$ in order to improve accuracy of the energy estimates. }
    \label{tab:hyperparameters}
\end{table}

\begin{table}[t]
  \centering
  \caption{Total energies and many-body charge gap $E_c=E(N{+}1)+E(N{-}1)-2E(N)$ of WSe$_2$/WS$_2$ at filling $\nu=2$ ($N=18$ holes in a $3\times3$ moir\'e supercell, lattice mismatch $\delta=0.02$), from neural-network variational Monte Carlo at the $27\,000$-step checkpoint. Uncertainties are Monte-Carlo standard errors computed from $16,384,000$ local energies sampled from the respective wavefunction. The standard error on $E_c$ is $[\rm{SE}_{N-1}^{2}+\rm{SE}_{N+1}^{2}+4\,\rm{SE}_{N}^{2}]^{1/2}$. All energies in meV.}
  \label{tab:nnvmc_energies}
  \begin{tabular}{ccccc}
    \toprule
    $\theta$ (deg) & $E(N{-}1)$ & $E(N)$ & $E(N{+}1)$ & $E_c$ \\
    \midrule
    \multicolumn{5}{l}{\textit{Relaxed}} \\
    2.0 & $-1388.90 \pm 0.01$ & $-1496.94 \pm 0.01$ & $-1604.36 \pm 0.01$ & $0.602 \pm 0.021$ \\
    2.2 & $-1456.09 \pm 0.01$ & $-1569.69 \pm 0.01$ & $-1682.62 \pm 0.01$ & $0.664 \pm 0.032$ \\
    2.4 & $-1523.81 \pm 0.02$ & $-1642.96 \pm 0.01$ & $-1761.28 \pm 0.01$ & $0.839 \pm 0.034$ \\
    2.6 & $-1591.60 \pm 0.01$ & $-1716.07 \pm 0.02$ & $-1839.24 \pm 0.01$ & $1.298 \pm 0.045$ \\
    2.8 & $-1659.02 \pm 0.01$ & $-1788.52 \pm 0.02$ & $-1915.77 \pm 0.03$ & $2.250 \pm 0.048$ \\
    \midrule
    \multicolumn{5}{l}{\textit{Rigid}} \\
    2.0 & $-1363.92 \pm 0.01$ & $-1471.62 \pm 0.01$ & $-1578.93 \pm 0.01$ & $0.391 \pm 0.026$ \\
    2.2 & $-1435.86 \pm 0.01$ & $-1549.26 \pm 0.01$ & $-1661.91 \pm 0.01$ & $0.746 \pm 0.022$ \\
    2.4 & $-1507.61 \pm 0.01$ & $-1626.57 \pm 0.01$ & $-1744.55 \pm 0.01$ & $0.973 \pm 0.024$ \\
    2.6 & $-1578.56 \pm 0.01$ & $-1702.56 \pm 0.01$ & $-1825.73 \pm 0.01$ & $0.821 \pm 0.030$ \\
    2.8 & $-1648.37 \pm 0.01$ & $-1777.46 \pm 0.01$ & $-1905.18 \pm 0.02$ & $1.370 \pm 0.035$ \\
    \bottomrule
  \end{tabular}
\end{table}
 
\begin{table}[t]
  \centering
  \caption{Charge gap $E_c$ averaged over the $25$--$27\,\mathrm{k}$ checkpoints with combined uncertainty $[\langle SE_{\infty}^{2}\rangle+s_{\mathrm{checkpoint}}^{2}]^{1/2}$, compared with the single-particle indirect band gap $\Delta_{12}$ on a converged $12\times12$ momentum mesh. $E_c^{\mathrm{corr}}=E_c-\Delta_{\mathrm{mesh}}$ removes the momentum-quantisation error of the $3\times3$ supercell, $\Delta_{\mathrm{mesh}}=\Delta_{12}^{3\times3}-\Delta_{12}$. All energies in meV.}
  \label{tab:charge_gap_lowtheta}
  \begin{tabular}{cccccccc}
    \toprule
    & \multicolumn{3}{c}{Relaxed} & & \multicolumn{3}{c}{Rigid} \\
    \cmidrule(lr){2-4}\cmidrule(lr){6-8}
    $\theta$ (deg) & $E_c$ & $\Delta_{12}$ & $E_c^{\mathrm{corr}}$ & & $E_c$ & $\Delta_{12}$ & $E_c^{\mathrm{corr}}$ \\
    \midrule
    2.0 & $0.62 \pm 0.03$ & 7.70 & 0.03 &  & $0.37 \pm 0.05$ & 1.92 & 0.37 \\
    2.2 & $0.69 \pm 0.05$ & 6.53 & 0.01 &  & $0.72 \pm 0.04$ & 1.91 & 0.72 \\
    2.4 & $0.86 \pm 0.06$ & 5.38 & -0.01 &  & $0.84 \pm 0.14$ & 1.69 & 0.64 \\
    2.6 & $1.26 \pm 0.07$ & 4.24 & 0.11 &  & $0.79 \pm 0.05$ & 1.02 & -0.07 \\
    2.8 & $2.20 \pm 0.07$ & 3.09 & 0.65 &  & $1.30 \pm 0.07$ & 0.23 & -0.34 \\
    \bottomrule
  \end{tabular}
\end{table}

\begin{table}[t]
  \caption{Many-body charge gap $E_c=E(N{+}1)+E(N{-}1)-2E(N)$ at filling $\nu=2$ ($N=18$ holes in a $3\times3$ moir\'e supercell) of WSe$_2$/WS$_2$ at lattice mismatch $\delta=0.02$, from neural-network variational Monte Carlo, compared with the single-particle indirect band gap $\Delta_{12}$ evaluated on a converged $12\times12$ momentum mesh. $E_c^{\mathrm{corr}}=E_c-\Delta_{\mathrm{mesh}}$ removes the momentum-quantisation error of the $3\times3$ supercell, $\Delta_{\mathrm{mesh}}=\Delta_{12}^{3\times3}-\Delta_{12}$. Uncertainties are $1\sigma$, $\sqrt{\langle SE_{\infty}^{2}\rangle+s_{\mathrm{checkpoint}}^{2}}$. All energies in meV.}
  \label{tab:charge_gap_lowtheta_full}
  \begin{ruledtabular}
  \begin{tabular}{ccccccc}
    & \multicolumn{3}{c}{Relaxed} & \multicolumn{3}{c}{Rigid} \\
    \cline{2-4}\cline{5-7}
    $\theta$ (deg) & $E_c$ & $\Delta_{12}$ & $E_c^{\mathrm{corr}}$ & $E_c$ & $\Delta_{12}$ & $E_c^{\mathrm{corr}}$ \\
    \hline
    2.0 & $0.97 \pm 0.04$ & 7.70 & 0.38 & $0.59 \pm 0.08$ & 1.92 & 0.59 \\
    2.2 & $1.08 \pm 0.08$ & 6.53 & 0.40 & $1.14 \pm 0.06$ & 1.91 & 1.14 \\
    2.4 & $1.35 \pm 0.10$ & 5.38 & 0.49 & $1.32 \pm 0.22$ & 1.69 & 1.12 \\
    2.6 & $1.98 \pm 0.11$ & 4.24 & 0.83 & $1.25 \pm 0.07$ & 1.02 & 0.38 \\
    2.8 & $3.46 \pm 0.11$ & 3.09 & 1.92 & $2.05 \pm 0.12$ & 0.23 & 0.40 \\
  \end{tabular}
  \end{ruledtabular}
\end{table}


\section{Supplemental results for 3R and 2H stacking}
\label{sec:results}

This section collects supplemental data for both stacking configurations: emergent gauge fields (Sec.~\ref{sec:gauge}), band structures at multiple twist angles (Sec.~\ref{sec:bands}), bandwidth and bandgap phase diagrams (Sec.~\ref{sec:bandwidth}), superfluid stiffness (Sec.~\ref{sec:stiffness}), and corresponding results for 2H (180$^\circ$) stacking (Sec.~\ref{sec:180}).

\subsection{Emergent gauge fields}
\label{sec:gauge}

\begin{figure*}
    \centering
    \includegraphics[width=0.7\linewidth]{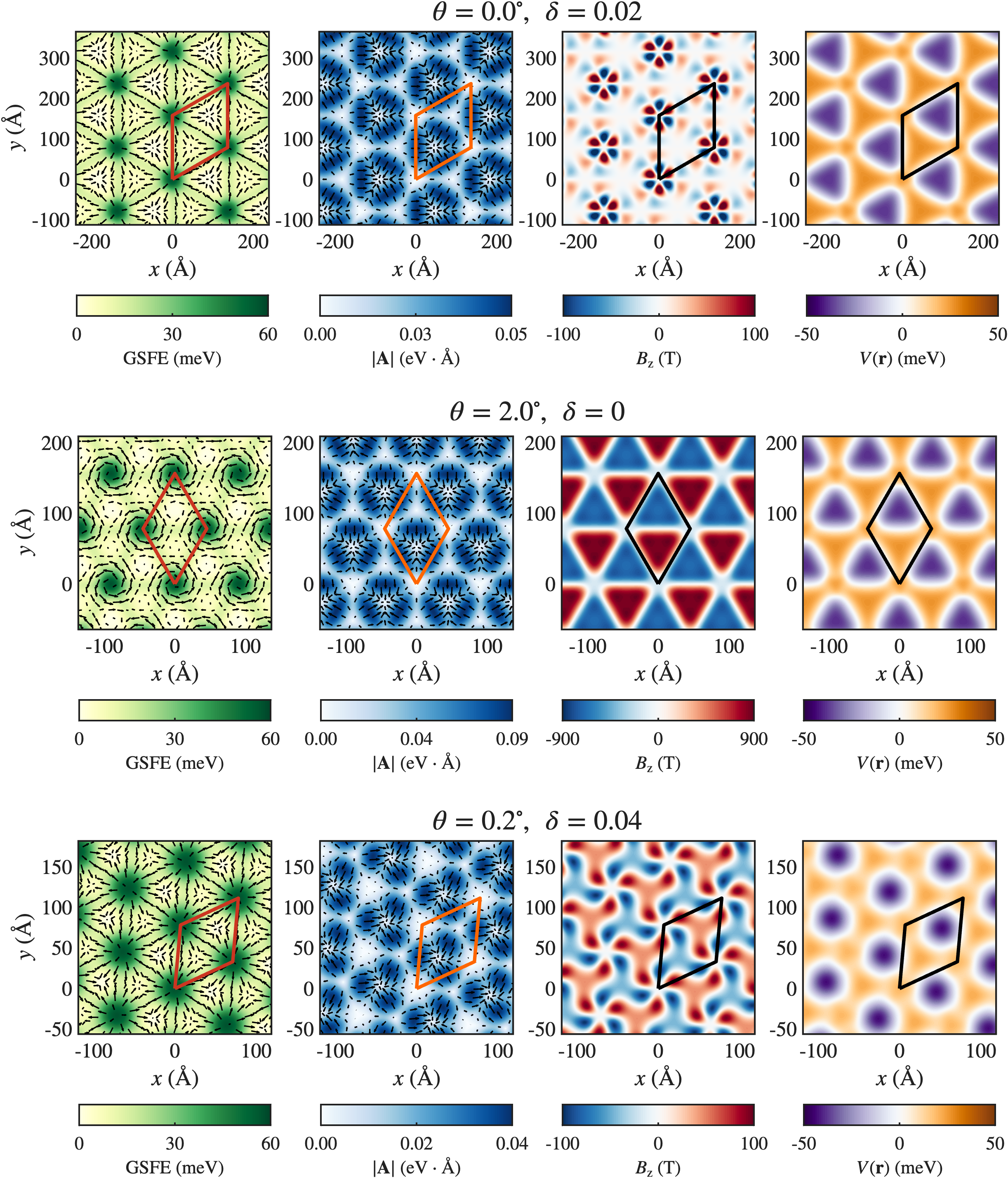}
    \caption{Emergent gauge fields from lattice relaxation in 3R-stacked (0$^\circ$) WSe$_2$/WS$_2$. Each row corresponds to a different point in the $(\theta, \delta)$ parameter space: (a) $\theta = 0.0^\circ$, $\delta = 0.02$; (b) $\theta = 2.0^\circ$, $\delta = 0$; (c) $\theta = 0.2^\circ$, $\delta = 0.04$. From left to right: GSFE with overlaid displacement vectors; $|\vec{A}|$ (eV$\cdot$\AA); $B_z$ (Tesla); and $V_{\mathrm{m}}(\vec{r})$ (meV). Colored parallelograms mark the moir\'e unit cell. The vector potential is plotted with the sign convention $\vec{A} = -\tfrac{1}{2}\mathcal{D}(\epsilon_1,\epsilon_2)$ [Eq.~\eqref{eqn:A_def}], matching the main text.}
    \label{fig:relax_0}
\end{figure*}

Figure~\ref{fig:relax_0} displays the relaxation-induced fields for 3R-stacked (0$^\circ$) WSe$_2$/WS$_2$ at three representative points in the $(\theta, \delta)$ parameter space. The leftmost column shows the GSFE landscape with overlaid displacement vectors, the second column shows the magnitude of the vector potential $|\vec{A}|$, the third column shows the pseudomagnetic field $B_z = \nabla \times \vec{A}$, and the rightmost column shows the relaxed moir\'e potential $V_{\mathrm{m}}(\vec{r})$.

The pseudomagnetic field $B_z$ exhibits an alternating-sign pattern, with regions of positive and negative effective field separated by domain walls. The field magnitude ranges from approximately 100~T for the mismatch-dominated case [$\theta = 0.0^\circ$, $\delta = 0.02$, Fig.~\ref{fig:relax_0}(a)] to approximately 900~T for the twist-dominated case [$\theta = 2.0^\circ$, $\delta = 0$, Fig.~\ref{fig:relax_0}(b)]. The experimentally relevant regime of WSe$_2$/WS$_2$ involves both twist angle and lattice mismatch ($\delta \approx 0.04$ for WSe$_2$/WS$_2$) simultaneously, as illustrated in Fig.~\ref{fig:relax_0}(c).

The vanishing of the net flux follows from the periodicity of the vector potential. By Stokes' theorem, the total flux is $\Phi = \oint_{\partial S} \vec{A} \cdot d\vec{l}$. Since $\vec{A}(\vec{r}) = \vec{A}(\vec{r} + \vec{R})$ for any moir\'e lattice vector $\vec{R}$, and the boundary consists of pairs of parallel edges traversed in opposite directions, all contributions cancel pairwise, giving $\Phi = 0$. The pseudo-$B$ field therefore cannot produce net Landau-level quantization. Instead, it creates a spatially modulated magnetic landscape that redistributes spectral weight and Berry curvature, driving topological transitions.

The pseudoelectric field $\Phi(\vec{r}) = (f_3 - f_4)\,\epsilon_0(\vec{r})$ is also concentrated at domain walls. For the mismatch-dominated case, the scalar strain reaches $\epsilon_0 \sim 1\%$, corresponding to a potential shift $|\Phi| \approx 20$~meV --- comparable to the moir\'e bandwidth at small angles. The two fields have complementary spatial distributions: $B_z$ is largest where the strain gradient is sharpest (at domain wall intersections), while $\Phi$ is largest where the dilation itself is maximal (along domain walls).

\subsection{Band structures at multiple twist angles}
\label{sec:bands}

\begin{figure}[h]
    \centering
    \includegraphics[width=0.7\linewidth]{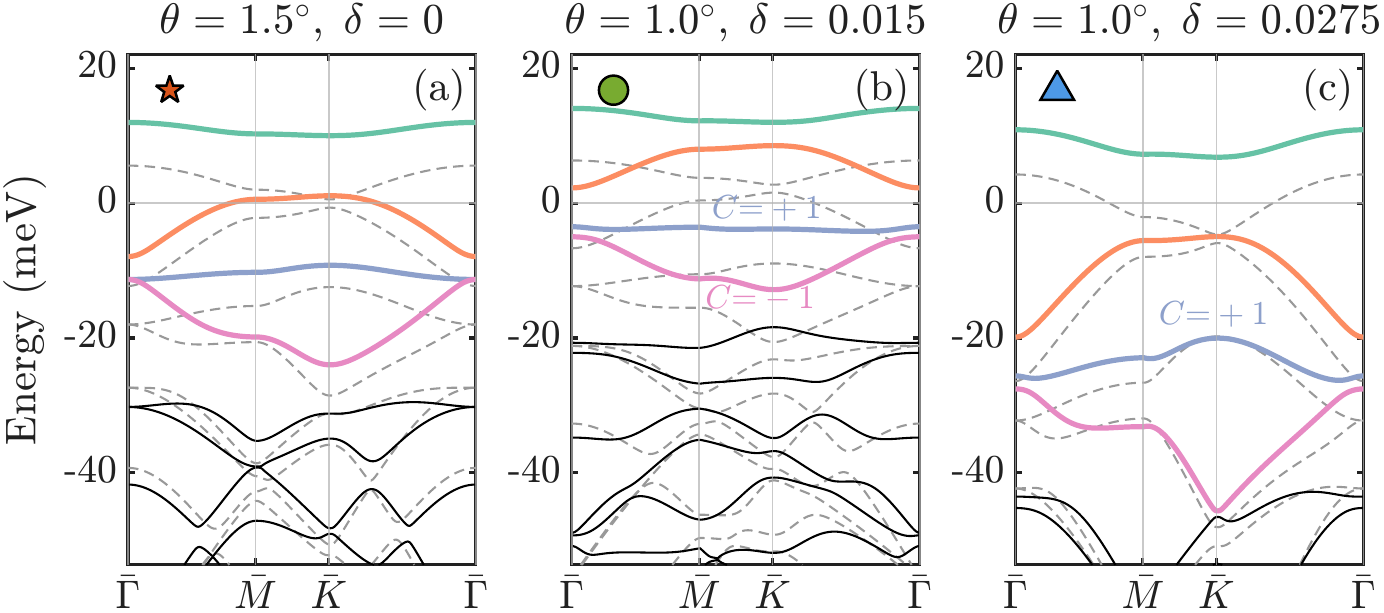}
    \caption{Band structures in 3R-stacked WSe$_2$/WS$_2$. Solid lines: relaxed; gray dashed: rigid. (a) $\theta = 1.5^\circ$, $\delta = 0$; (b) $\theta = 1.0^\circ$, $\delta = 0.015$; (c) $\theta = 1.0^\circ$, $\delta = 0.0275$. Chern numbers of bands with $C \neq 0$ are labeled.}
    \label{fig:bands_multi}
\end{figure}

Figure~\ref{fig:bands_multi} shows the rigid and relaxed band structures at three representative points in the $(\theta, \delta)$ parameter space. At $\theta = 1.5^\circ$, $\delta = 0$ [Fig.~\ref{fig:bands_multi}(a)], the topology is trivial. At $\theta = 1.0^\circ$, $\delta = 0.015$ [Fig.~\ref{fig:bands_multi}(b)], the third and fourth valence bands are separated by a clear gap $\Delta_{34}$ and carry compensating Chern numbers $C_3 = +1$, $C_4 = -1$. At $\theta = 1.0^\circ$, $\delta = 0.0275$ [Fig.~\ref{fig:bands_multi}(c)], the third valence band remains topological with $C = +1$. The topmost two valence bands remain trivial in all three cases, consistent with the decomposition in the main text.

\subsection{Chern number phase diagrams}
\label{sec:chern}

\begin{figure*}
    \centering
    \includegraphics[width=0.8\linewidth]{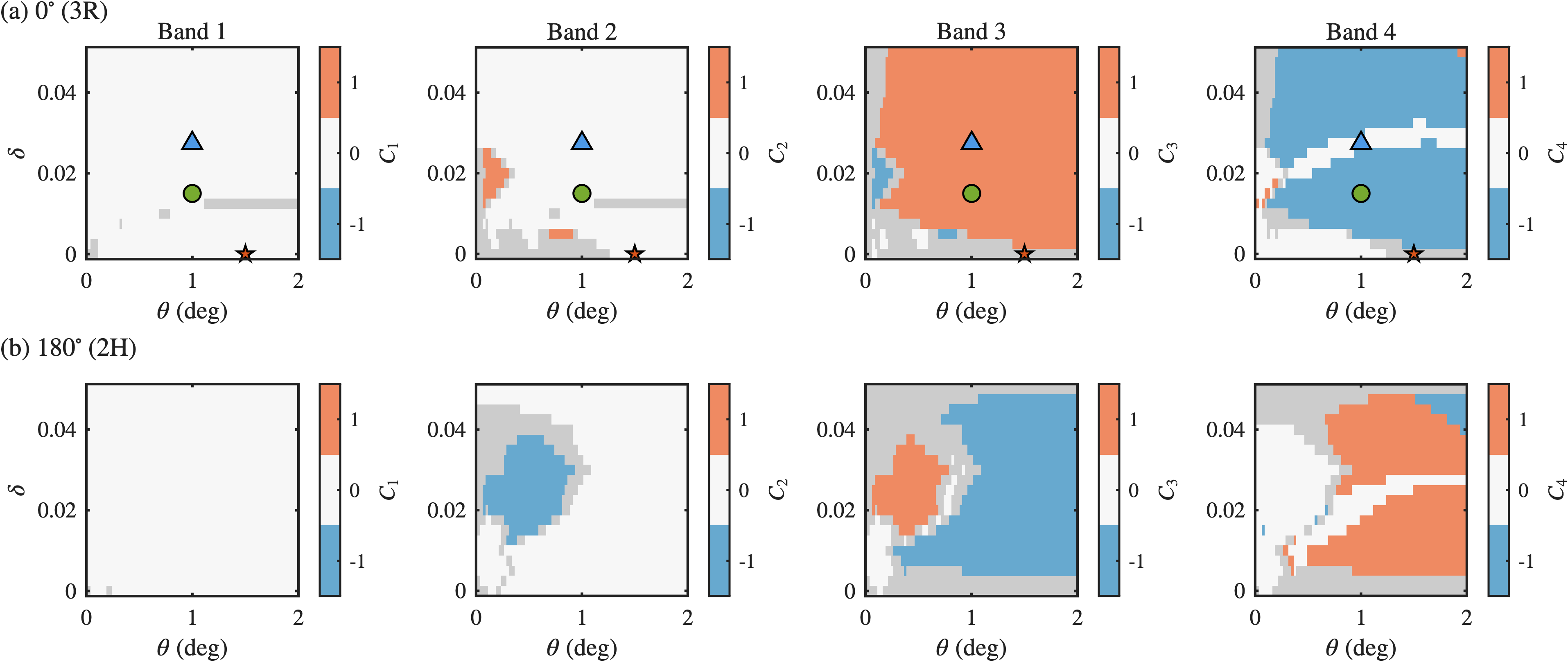}
    \caption{Chern number phase diagrams for (a) 3R-stacked (0$^\circ$) and (b) 2H-stacked (180$^\circ$) WSe$_2$/WS$_2$. Left to right: $C_1$, $C_2$, $C_3$, $C_4$. Gray regions: Chern number ill-defined (corresponding gap $< 0.8$~meV). In (a), the star, circle, and triangle symbols denote the parameters $(\theta, \delta)$ of Fig.~\ref{fig:bands_multi}(a)--(c), respectively.}
    \label{fig:topo_combined}
\end{figure*}

Figure~\ref{fig:topo_combined} shows the Chern numbers of the top valence bands as functions of $(\theta, \delta)$ for both stacking configurations. For 3R stacking [Fig.~\ref{fig:topo_combined}(a)], the first and second valence bands remain trivial ($C_1 = C_2 = 0$) over the entire parameter range, while the third and fourth valence bands acquire compensating Chern numbers $C_3 = +1$, $C_4 = -1$ over a broad region of small twist angles and finite lattice mismatch (corresponding to Fig.~3 of the main text). Gap-closure contours (solid: lower gap; dashed: upper gap) delineate the topological phase boundaries. For 2H stacking [Fig.~\ref{fig:topo_combined}(b)], the topological structure is markedly simpler: the nontrivial region is smaller and appears only at small $\theta$ and moderate $\delta$, reflecting the more symmetric GSFE landscape, which produces weaker domain-wall sharpening and correspondingly smaller strain-induced perturbations.

\subsection{Bandwidth and bandgap phase diagrams}
\label{sec:bandwidth}

\begin{figure*}
    \centering
    \includegraphics[width=\linewidth]{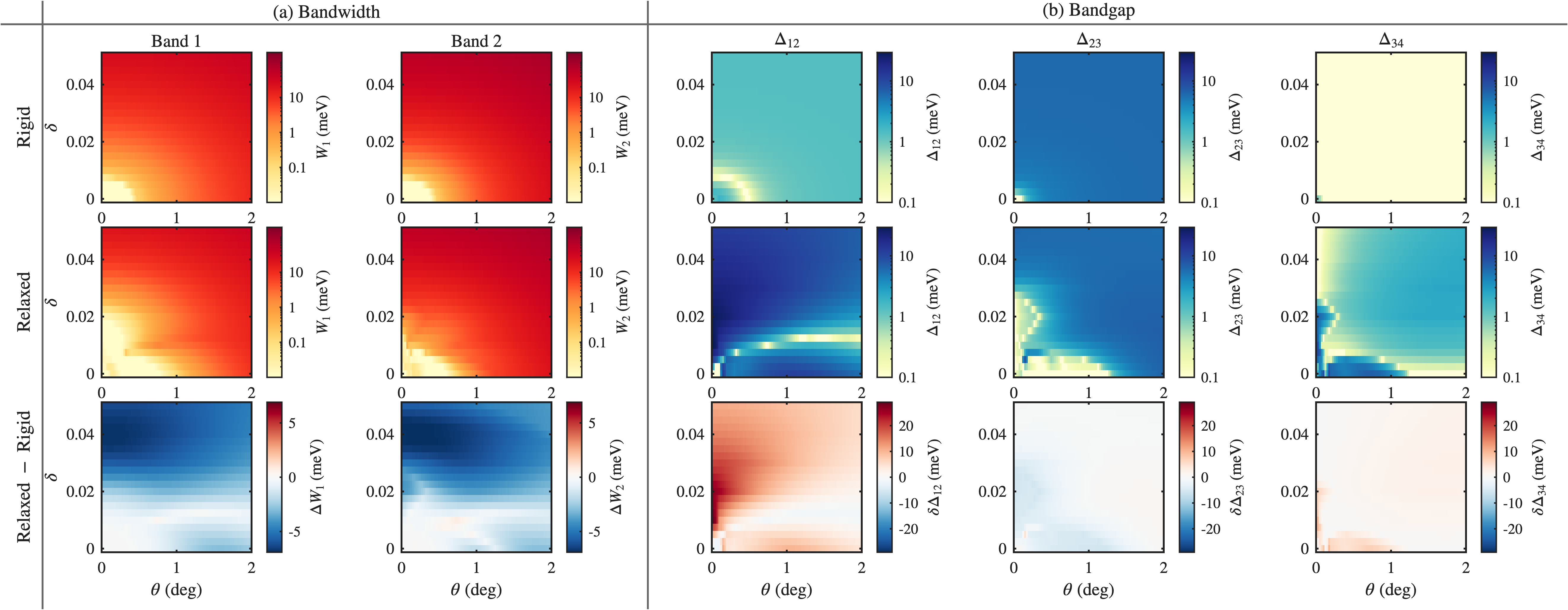}
    \caption{Bandwidth $W_n$ (left) and band gaps $\Delta_{12}$, $\Delta_{23}$, $\Delta_{34}$ (right) for 3R-stacked WSe$_2$/WS$_2$. Rows: rigid (top), relaxed (middle), difference (bottom). Bandwidths use a logarithmic color scale.}
    \label{fig:bandwidth_0}
\end{figure*}

Figure~\ref{fig:bandwidth_0} shows the bandwidths $W_n$ and remote gaps $\Delta_{12}$, $\Delta_{23}$, $\Delta_{34}$ as functions of $(\theta, \delta)$ for 3R stacking. At small twist angles ($\theta \lesssim 2^\circ$), relaxation narrows the bandwidths of the top valence bands, consistent with the sharpening of the moir\'e potential: the deeper potential wells produce more tightly confined states with reduced dispersion. The bandwidth reduction is most pronounced in the flattest regions of the phase diagram, where the rigid-model bandwidth is already below 10~meV and relaxation further suppresses it by a factor of 2-3. Relaxation enhances the gap $\Delta_{12}$ between the first and second bands across most of the parameter range, further isolating the topmost band. Relaxation also enhances the gap $\Delta_{34}$ between the third and fourth valence bands: although the pseudoelectric and higher-harmonic moir\'e channels acting alone narrow $\Delta_{34}$, the pseudomagnetic field more than compensates by opening a topological gap. The zero-contour of $\Delta_{34}$ within the relaxed model coincides with the topological phase boundary of Fig.~\ref{fig:topo_combined}(a).

\subsection{Superfluid stiffness phase diagrams}
\label{sec:stiffness}

The superfluid weight $D_s$ of a flat-band superconductor is controlled by the integrated quantum metric~\cite{peotta2015superfluidity, torma2022superconductivity}, and in two dimensions $D_s$ determines the BKT transition temperature $T_\mathrm{BKT} \propto D_s$. We compute $\bar{D}_s$ as the mean eigenvalue of the $2\times2$ superfluid weight tensor $D_{s,\a\b} = |\Gamma_\mathrm{m}^*|^{-1}\!\int_\mathrm{mBZ} g_{\a\b}(\vec{k})\, d^2k$.

Figure~\ref{fig:geometry_0} shows $\bar{D}_s$ for the top valence bands in 3R-stacked WSe$_2$/WS$_2$, including the topological third valence band. Relaxation generically reduces $\bar{D}_s$ over much of the phase diagram --- by more than 50\% in some regions --- reflecting the smoothing of the quantum metric. The reduction is most pronounced at small twist angles where relaxation is strongest and the rigid-model metric has the sharpest peaks.

For a band with Chern number $C$, the integrated metric satisfies $\int g_\mathrm{Tr}\, d^2k \geq 2\pi |C|$~\cite{peotta2015superfluidity}, with saturation defining the ideal-Chern (K\"ahler) limit. The integrated trace-condition violation $T/2\pi = (2\pi)^{-1}\!\int(\mathrm{Tr}\,g - |\Omega|)\,d^2k$ and Berry-curvature fluctuation $F$, which together characterize this proximity, are mapped over $(\theta, \delta)$ in Fig.~4 of the main text. The smoothing of quantum geometry by relaxation thus has competing implications for correlated phases: the reduced total metric lowers the geometric weight available to stabilize fractional Chern insulators, while the improved uniformity moves the band closer to ideal conditions.

\begin{figure}
    \centering
    \includegraphics[width=0.5\linewidth]{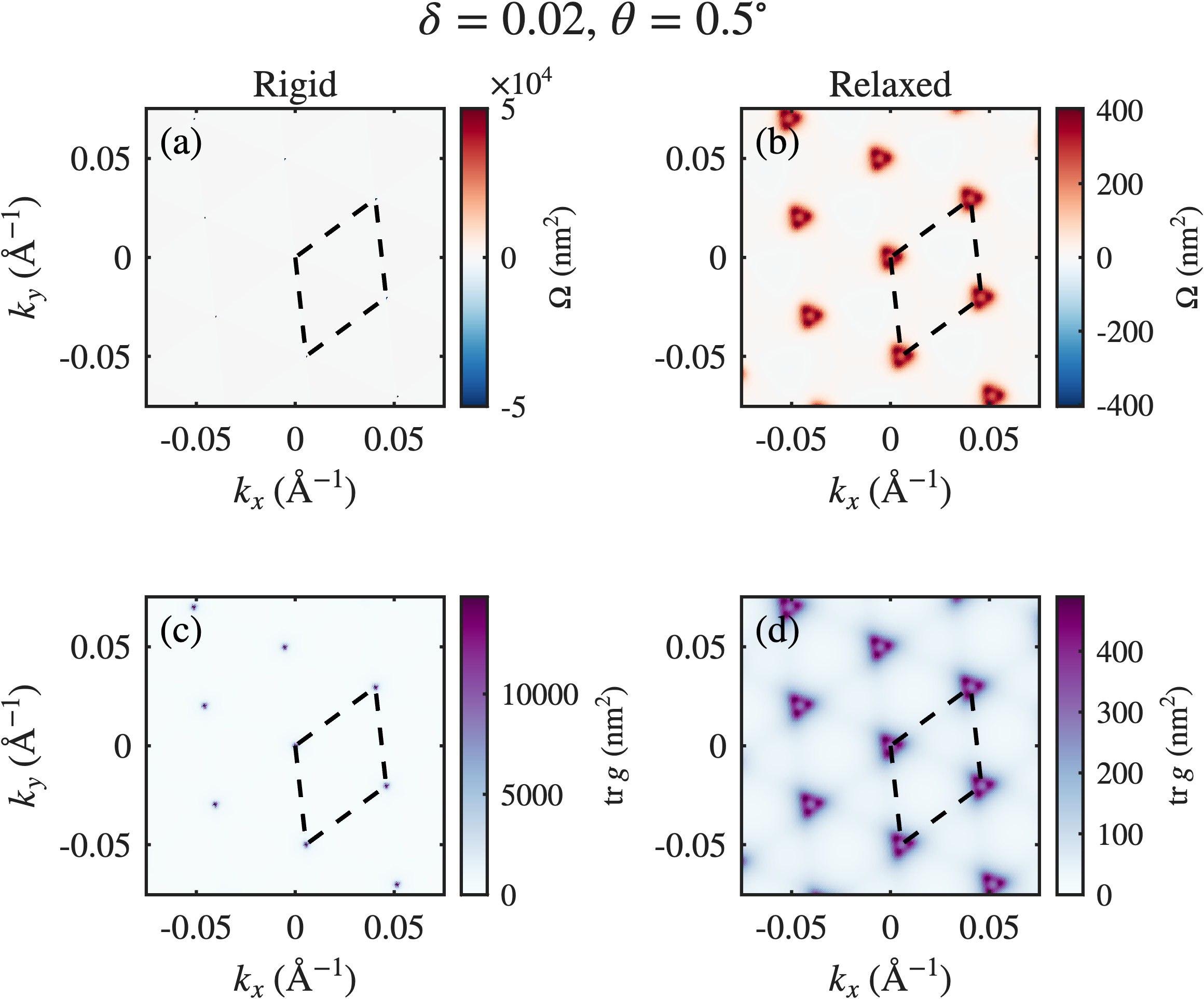}
    \caption{Berry curvature $\Omega(\bm{q})$ and trace of the quantum metric $\mathrm{tr}\,g(\bm{k})$ of the third valence band of WSe$_2$/WS$_2$ at $\delta = 0.02$, $\theta = 0.5^\circ$, comparing the rigid (left column) and relaxed (right column) models. (a),(b)~$\Omega(\bm{q})$. (c),(d)~$\mathrm{tr}\,g(\bm{q})$. Dashed rhombi mark the moir\'e Brillouin zone. In the rigid limit, both quantities are sharply concentrated at the $\bar{K}$ corners with peak values exceeding $10^4$~nm$^2$, whereas relaxation redistributes them smoothly across the Brillouin zone, with peak values reduced by roughly two orders of magnitude. Note the different color scales between rigid and relaxed panels.}
    \label{fig:berry}
\end{figure}

\begin{figure}[h]
    \centering
    \includegraphics[width=0.6\linewidth]{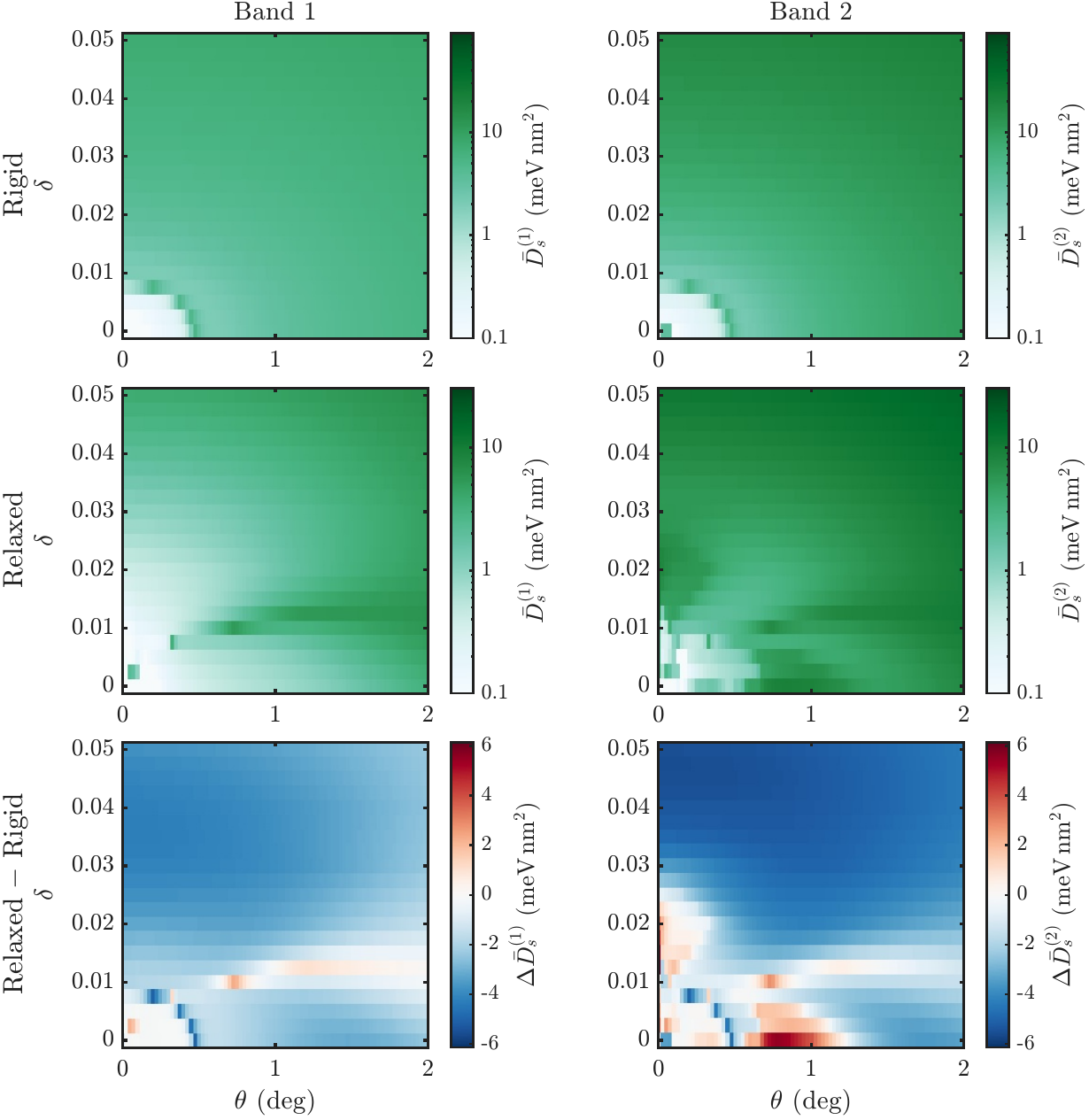}
    \caption{Superfluid stiffness $\bar{D}_s$ for the top valence bands of 3R-stacked WSe$_2$/WS$_2$: rigid (top), relaxed (middle), and difference (bottom). Contours mark gap closures.}
    \label{fig:geometry_0}
\end{figure}

The momentum-space distributions underlying these integrated diagnostics are shown in Fig.~\ref{fig:berry} for a representative point ($\delta = 0.02$, $\theta = 0.5^\circ$). In the rigid model, the Berry curvature $\Omega(\bm{k})$ and quantum metric trace $\mathrm{tr}\,g(\bm{k})$ are both sharply localized at the $\bar{K}$ corners of the moir\'e Brillouin zone, with peak values exceeding $10^4$~nm$^2$. Lattice relaxation redistributes both quantities over the Brillouin zone—reducing peak values by roughly two orders of magnitude while preserving the integrated Chern number $C_3 = +1$. The Berry curvature retains residual concentration near $\bar{K}$, but the trace of the metric becomes nearly uniform, which is the real-space origin of the reduced trace-condition violation $T/2\pi$ reported in Fig.~4 of the main text.

\subsection{Results for 2H (180$^\circ$) stacking}
\label{sec:180}

We present the corresponding results for 2H-stacked (180$^\circ$) WSe$_2$/WS$_2$. The 2H stacking configuration differs from 3R in the symmetry of the GSFE landscape: the antisymmetric Fourier coefficients (the $c_4$ and $c_5$ terms in the GSFE expansion) are substantially smaller in magnitude for 2H ($c_4 = +0.70$~meV) than for 3R ($c_4 = -3.68$~meV), as shown in Fig.~\ref{fig:gsfe}. This produces a GSFE landscape with higher effective symmetry between the two triangular domain types. Below we show the emergent gauge fields, Chern number phase diagrams, bandwidth and bandgap phase diagrams, and quantum geometry for 2H stacking, paralleling the 3R results presented above.

\subsubsection{Emergent gauge fields}

\begin{figure*}
    \centering
    \includegraphics[width=0.7\linewidth]{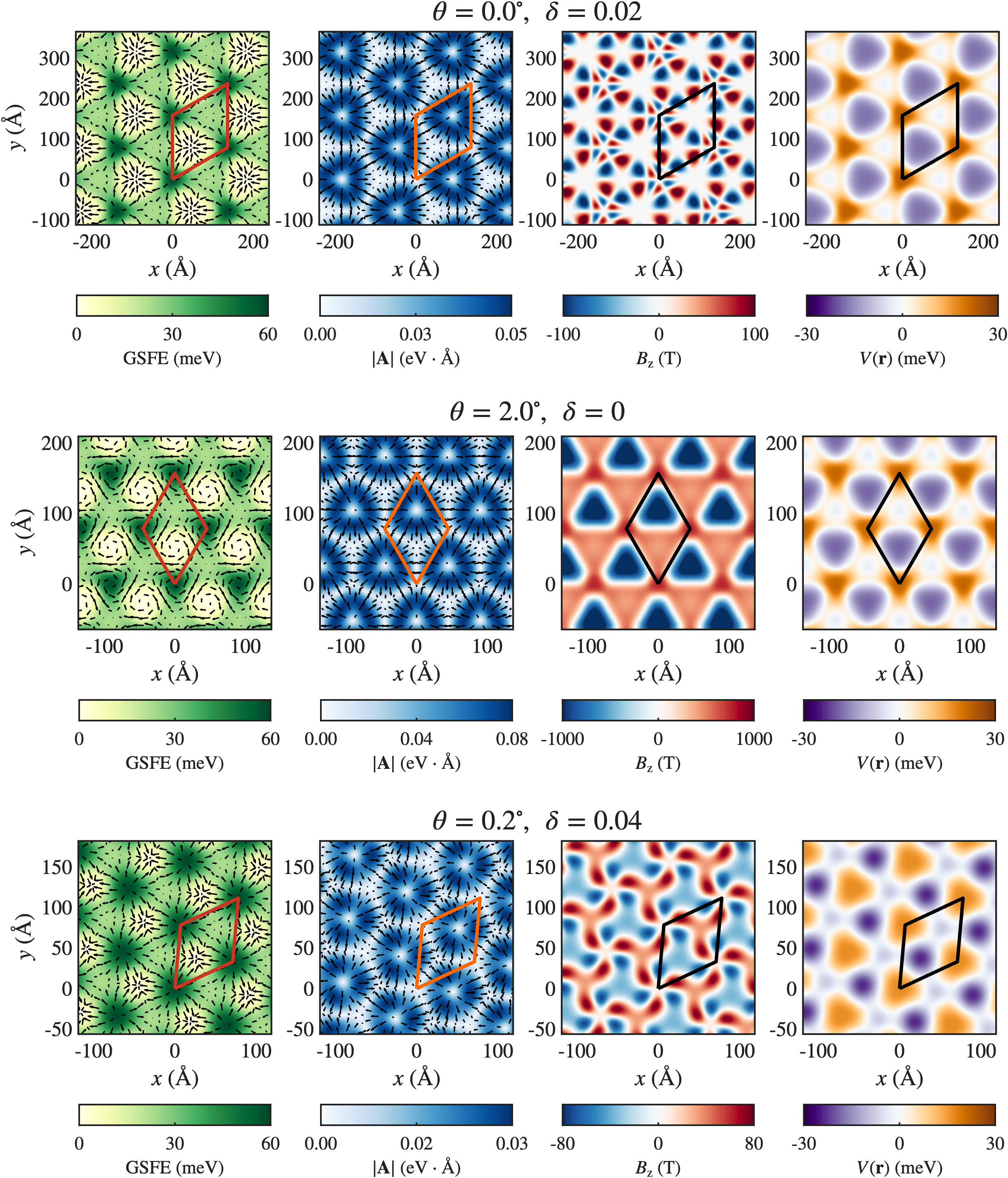}
    \caption{Same as Fig.~\ref{fig:relax_0} but for 2H-stacked (180$^\circ$) WSe$_2$/WS$_2$.}
    \label{fig:relax_180}
\end{figure*}

Figure~\ref{fig:relax_180} shows the relaxation-induced fields for 2H stacking at the same three $(\theta, \delta)$ points as Fig.~\ref{fig:relax_0}. The displacement vectors, vector potential, pseudomagnetic field, and moir\'e potential all display the expected $C_3$ symmetry, but the spatial pattern within the moir\'e unit cell differs from the 3R case. In 2H stacking, the reduced asymmetry between the two domain types produces a more uniform relaxation pattern with less pronounced domain walls compared to 3R at the same moir\'e wavelength. The pseudomagnetic field $B_z$ reaches comparable peak magnitudes (hundreds of Tesla, up to ${\sim}1000$~T for the twist-dominated case), but the spatial distribution is rearranged according to the different high-symmetry points of the 2H GSFE landscape.

\subsubsection{Chern number phase diagram}

The Chern number phase diagram for 2H stacking is shown in Fig.~\ref{fig:topo_combined}(b). In contrast to 3R stacking, the topological region between the third and fourth valence bands is confined to a smaller portion of the $(\theta, \delta)$ plane, and the first and second valence bands remain trivial throughout.

\subsubsection{Bandwidth and bandgap phase diagrams}

\begin{figure*}
    \centering
    \includegraphics[width=\linewidth]{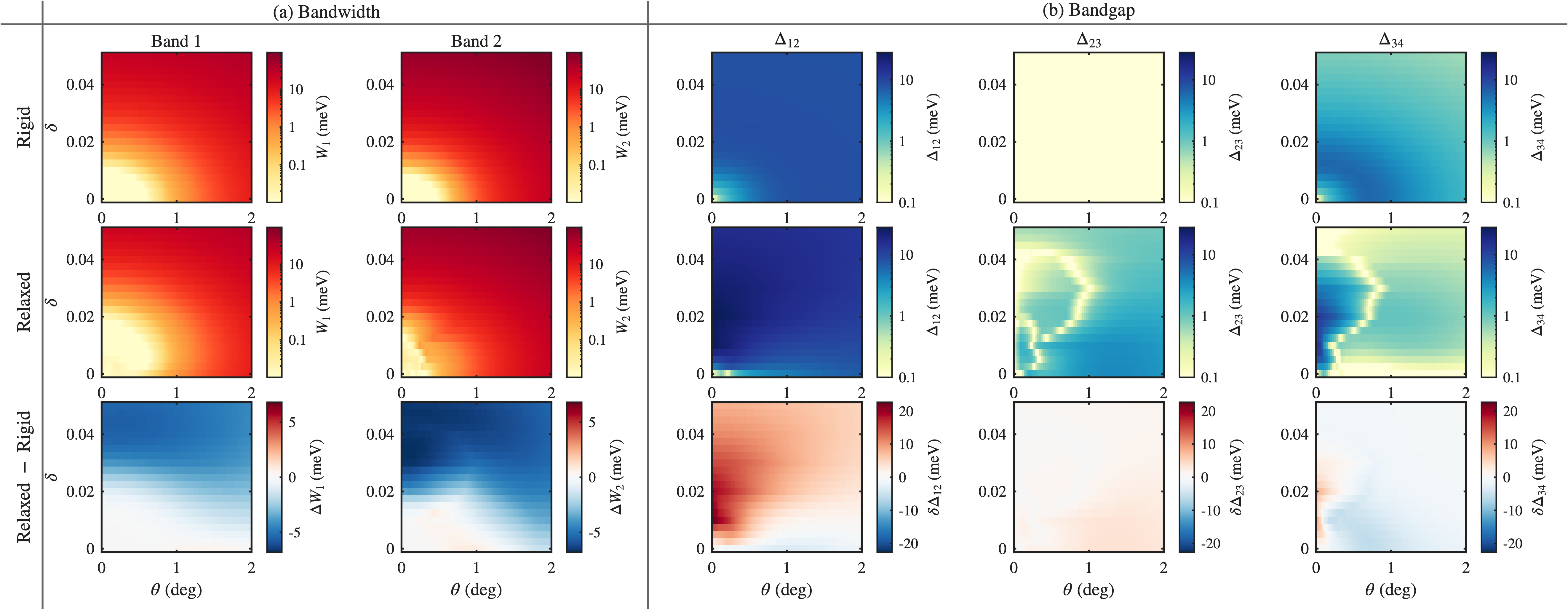}
    \caption{Same as Fig.~\ref{fig:bandwidth_0} but for 2H-stacked WSe$_2$/WS$_2$.}
    \label{fig:bandwidth_180}
\end{figure*}

Figure~\ref{fig:bandwidth_180} shows the bandwidth and remote gap phase diagrams for 2H stacking. The qualitative trends are similar to the 3R case: relaxation narrows the bandwidths at small twist angles and enhances both $\Delta_{12}$ and $\Delta_{34}$. The bandwidth reduction is somewhat less pronounced in 2H than in 3R, reflecting the weaker domain-wall sharpening associated with the more symmetric GSFE.

\subsubsection{Quantum geometry}

\begin{figure}[h]
    \centering
    \includegraphics[width=0.6\linewidth]{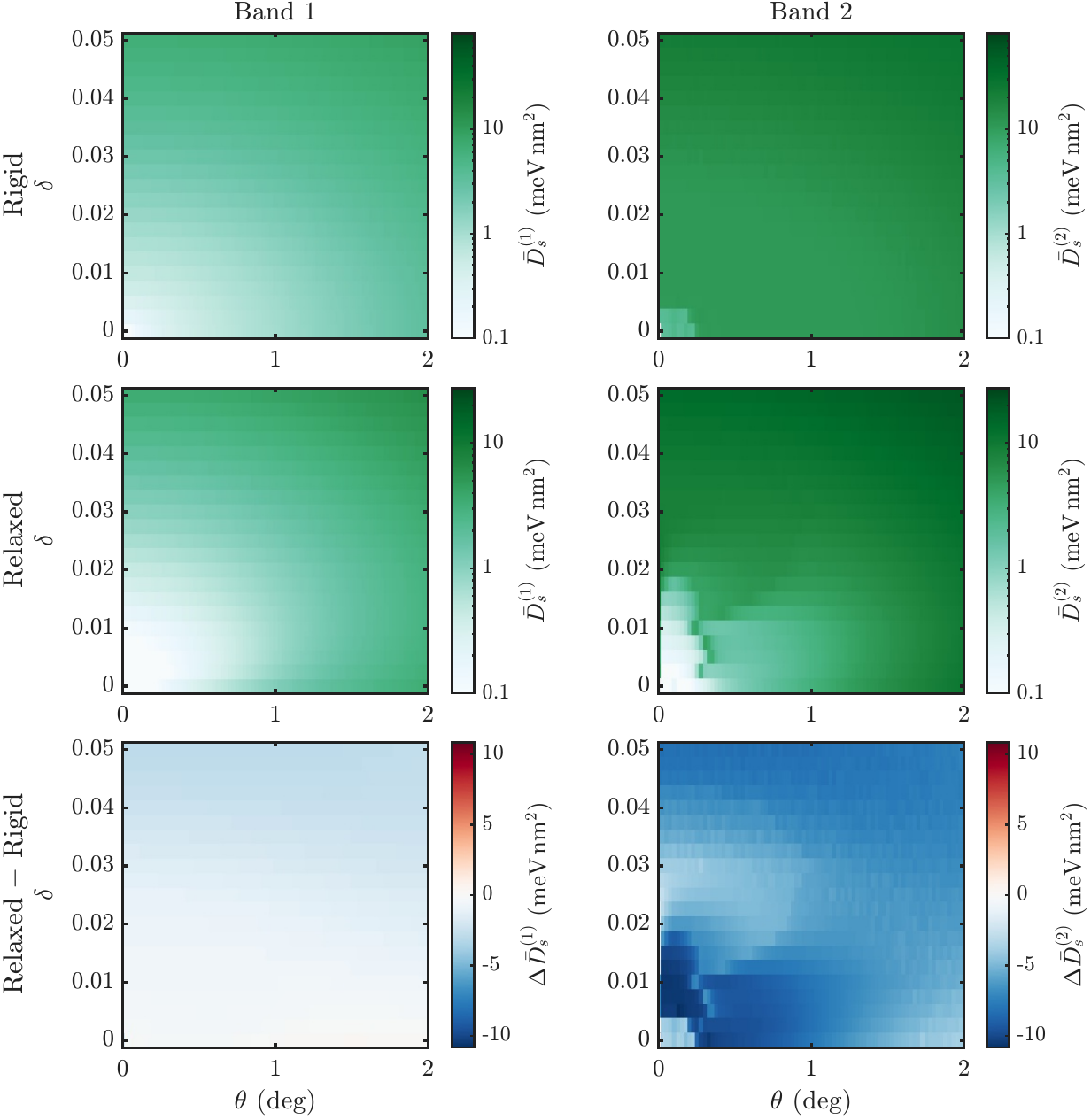}
    \caption{Same as Fig.~\ref{fig:geometry_0} but for 2H-stacked WSe$_2$/WS$_2$.}
    \label{fig:geometry_180}
\end{figure}

Figure~\ref{fig:geometry_180} shows the superfluid stiffness $\bar{D}_s$ for 2H stacking. As in the 3R case, relaxation generically reduces $\bar{D}_s$ relative to the rigid model. The reduction is most significant at small twist angles where relaxation is strongest. For 2H stacking, the rigid-model $\bar{D}_s$ values are generally smaller than for 3R at the same $(\theta, \delta)$, reflecting the different band structure and inter-band coupling landscape. The difference between relaxed and rigid $\bar{D}_s$ is correspondingly smaller in absolute terms, though the fractional reduction remains comparable.

\clearpage
\bibliography{ref}